\documentclass{article}
\usepackage[utf8]{inputenc}
\usepackage{authblk}
\usepackage{hyperref}
\usepackage{bm}
\usepackage{amsmath}

\title {From Drag to Invariance: The Experimental Pressure Behind Special Relativity.}
\author{Galina Weinstein\thanks{The Department of Philosophy, University of Haifa.}}

\begin{document}

\maketitle

\begin{abstract}
This paper completes a three-part study of Einstein’s 1905 special relativity by reconstructing the experimental pressures that shaped his thinking from 1895 to June 1905. Following Stachel’s historiographical line, I trace Einstein’s path under the cumulative weight of a series of recalcitrant experiments: stellar aberration, Arago’s prism test, Fresnel’s partial-drag account, Boscovich’s proposal of a water-filled telescope, Fizeau’s water-tube result, ether-drift null experiments, and the magnet–conductor induction asymmetry. Lorentz’s electron theory attempted to domesticate these findings within a fragile ether framework, while Einstein—still loyal to Maxwell’s equations—became increasingly troubled by their conflict with Galilean kinematics. The paper examines Einstein’s temporary adherence to emission theory and its decisive breakdown in light of Fizeau’s result. In this reconstruction, the 1905 paper does not emerge as a kinematic postulate ex nihilo, but as a principled resolution forced by an interconnected complex of experimental anomalies.
\end{abstract}

\section{Introduction}

This paper completes a three-part study of Einstein’s 1905 special relativity by reconstructing his development from 1895 to June 1905 under the pressure of decisive experiments. I retrace Einstein’s approach to relativity through the empirical tensions that repeatedly undermined ether-based electrodynamics: stellar aberration, Arago’s prism test, Fresnel’s partial-drag hypothesis, Boscovich’s proposal of a water-filled telescope, Fizeau’s water-tube result, ether-drift null experiments, and the magnet–conductor asymmetry.

The narrative begins with a historical and analytical reconsideration of aberration and its successive treatments, running from Arago through Fresnel and culminating in Fizeau, whose result rendered Fresnel’s drag coefficient \emph{de facto} canonical and posed a fatal difficulty for emission theory, to which Einstein had temporarily been inclined. By the end of the nineteenth century, the “clouds” had accumulated. Lorentz’s electron theory was an ambitious but conceptually fragile construction designed to reconcile these findings within a delicately balanced ether framework. Einstein engaged directly with both the experimental situation and Lorentz’s 1895 \emph{Versuch}. Yet, he remained a committed adherent of Maxwell’s electrodynamics and was deeply troubled by its \emph{mutatis mutandis} conflict with Newtonian mechanics.

In the reconstruction offered here, Einstein’s 1905 paper emerges not as a kinematic postulate \emph{ex nihilo}, but as a principled resolution to an intertwined cluster of experimental anomalies and theoretical contradictions. If disentangling the historical threads resembles working at a Gordian knot \cite{Stachel2002}, the following argument suggests that Einstein cut through it not by invention in a vacuum, but by sustained confrontation with experiment.

\section{The Clouds of Fin-de-Siècle Physics}

\subsection{Bradley and Aberration} \label{BA}

James Bradley announced the discovery of stellar aberration in his 1728 communication to the Royal Society (in a letter to Edmond Halley published in 1729). His original campaign was designed not to detect aberration but to detect annual stellar parallax, thereby providing direct empirical confirmation of the Earth’s orbital motion. To this end, he selected stars passing close to the zenith (notably $\gamma$ Draconis) to minimize refraction and maximize parallactic displacement.

After roughly one year of zenith-sector observations of $\gamma$ Draconis, Bradley detected a strictly periodic annual displacement of order $\sim20''$. By March 1726, the star lay about $20''$ farther south than at the start of the series, had returned by June to its December position, and by September lay about $20''$ north, completing a full twelve-month cycle. This variation's temporal phase and symmetry made it incompatible with stellar parallax.

Bradley then systematically eliminated instrumental causes. He replaced and tested the plumb line, verified the rigidity of the arc and the position of the focus wire, and compared stars nearly opposite in right ascension; the residual pattern remained unchanged, establishing that the instrument did not produce the phenomenon. He next considered whether a nutation of the Earth’s axis could account for the effect, but rejected this based on contrary variations in comparison stars \cite{Brad}.

With instrumental and terrestrial explanations ruled out, Bradley identified a new empirical regularity. The departures in declination varied in proportion to the versed sine of the Sun’s distance from the equinoxes, indicating a dependence on the Sun’s position rather than stellar geometry. After completing a more precise zenith sector at Wanstead in 1727, he further found that stars reached their extrema when transiting the meridian at six o’clock (either morning or evening), revealing that the extrema were keyed to the relative orientation of the Earth’s orbital motion rather than to any intrinsic property of the observed stars.

At this stage, Bradley identified the actual cause of the phenomenon as the joint effect of the finite (progressive) speed of light and the Earth’s annual orbital motion. Since the observer (the eye in the tube, i.e., telescope) is in motion while the light is in transit, the true stellar and observed directions cannot coincide. The star will be seen in the direction in which the tube is actually pointed, not in its true geometrical position; consequently, the tube must be inclined slightly forward (relative to the true stellar direction) for the light to enter it while the observer is moving \cite{Brad}.%
\footnote{“Although therefore the true or real Place of an Object is perpendicular to the Line in which the Eye is moving, yet the visible Place will not be so, since that, no doubt, must be in the Direction of the Tube; but the Difference between the true and apparent Place will be (\emph{cateris paribus}) greater or less, according to the different Proportion between the Velocity of Light and that of the Eye” \cite{Brad}.}

From the geometrical construction, he derived a law relating the apparent displacement to the ratio of the observer’s velocity to the velocity of light, showing that the sine of the difference between the true and visible direction is proportional to that velocity ratio.
Taking the Earth’s orbital speed to be approximately one thousandth of the speed of light, Bradley obtained a predicted displacement of about $20''$, in close agreement with the observations. He further deduced the form of the apparent stellar paths implied by the hypothesis: a star at the ecliptic pole should describe an apparent circle of roughly $20''$ radius, whereas stars at other ecliptic latitudes should trace ellipses of correspondingly reduced semi-axes—again in accordance with observation.

At this point, Bradley shifted from explanation to measurement, announcing that he would use the data to determine the ratio of the velocity of light to the Earth’s orbital velocity on the assumption that the observed phenomena were produced by the causes he had identified.

From the observed amplitude ($\approx 20.5''$) and the known orbital velocity of the Earth, he derived a speed of light consistent with Ole Rømer's determinations from Jupiter’s satellites. He observed the same displacement across stars of different magnitudes and presumed distances, thereby confirming the distance-independence predicted by the aberration hypothesis \cite{Brad}.

Bradley’s result supplied a new dynamical argument for the Earth’s orbital motion, independent of the long-sought stellar parallax. It simultaneously provided a further empirical confirmation of the finite speed of light (light does not arrive instantaneously). The same coherent law explains the annual displacement only on the joint assumption that the Earth moves and that light propagates with finite velocity; denying one requires denying the other. Once recognized, it immediately carried Copernican force. Thus, Bradley closes his letter to Halley by noting that \cite{Brad}:

\begin{quote}
There appearing therefore after all, no sensible Parallax in the fixed Stars, the \emph{Anti-Copernicans} have still room on that Account, to object against the Motion of the Earth; and they may have (if they please) a much greater Objection against the Hypothesis, by which I have endeavoured to solve the fore-mentioned \emph{Phenomena}; by denying the progressive Motion of Light, as well as that of the Earth. But as I do not apprehend, that either of these Postulates will be denied me by the Generality of the Astronomers and Philosophers of the present Age; so I shall not doubt of obtaining their Assent to the Consequences, which I have deduced from them; if they are such as have the Approbation of so great a Judge of them as yourself.    
\end{quote}
The Anti-Copernicans rested their objection on the absence of observed parallax. If the Earth moves, parallax must appear; since none is seen, the Earth does not move. Bradley acknowledges that this objection remains formally in place due to the continued absence of parallax. But to dismiss aberration as evidence of Earth’s motion, one must also deny the finite speed of light, a premise already accepted by virtually all contemporary astronomers and philosophers. In 1729, that denial was no longer intellectually credible.

\subsection{The Corpuscular Theory of Light} 

At the beginning of the nineteenth century, two rival conceptions of light dominated scientific debate: the Newtonian corpuscular theory, which treated light as a stream of particles, and the wave theory, which regarded it as a disturbance propagated through an ether. Each framework sought to account for the growing range of optical phenomena and, in doing so, to strengthen its claim to validity.

In the Newtonian corpuscular theory, the corpuscle’s total velocity increases in the denser medium, but not equally in all directions.
Two velocity components are involved when light passes from one medium to another: 

\noindent 1. The tangential component $v_t$ is parallel to the interface (unchanged since there is no force along the surface). 

\noindent 2. The normal component $v_n$ is perpendicular to the interface (increased by the attractive force exerted by the denser medium).

Faster motion in the denser medium produces refraction toward the normal. 
Before entering the medium (say, from air), the light corpuscle travels with components $(v_{t1}, v_{n1})$.  
After entering the denser medium (say, water), the tangential velocity remains the same ($v_{t2} = v_{t1}$), while the normal velocity increases ($v_{n2} > v_{n1}$) due to the attractive force.  The direction of motion, that is, the angle of refraction $r$, depends on the ratio of these components:
\begin{equation} \label{ratio}
\tan r = \frac{v_t}{v_n}.    
\end{equation}
If $v_n$ increases while $v_t$ remains the same, $\tan r$ decreases and $r$ becomes smaller; in other words, the ray bends toward the normal.  

\subsection{Arago's Prism Experiment}

In 1810, François Arago—working within the framework of the corpuscular theory of light—proposed an aberration experiment.
Arago used Bradley’s result as the accepted starting point, and then asked whether further effects depending on the velocity of light can be detected (in particular via refraction). He proposed to test whether light particles entering a prism would be refracted differently according to their velocity relative to the prism \cite{Arago}. 

If the Earth moves through space with velocity $v_{\text{Earth}}$, then the total velocity of the light corpuscle relative to the prism $v'$ should be modified according to the Newtonian law of velocity addition:

\begin{equation} \label{vlight}
v' = v_{\text{light}} \pm v_{\text{Earth}}.
\end{equation}
Here, $v_{\text{light}}$ is the intrinsic velocity of the light particle with respect to the source (the star). 

\noindent $v_{\text{Earth}}$ is the velocity of the Earth (and hence of the prism) relative to the star.

\noindent The $\pm$ sign depends on whether Earth is moving toward or away from the source.

\noindent $v'$ is the resultant velocity magnitude of the corpuscle as seen from the moving prism frame. 

$v_t$ and $v_n$ are the components of the light corpuscle’s velocity in the prism’s rest frame. Depending on whether the Earth moves toward or away from the star, $v_t$ and $v_n$ of the corpuscle with respect to the prism change slightly.  
Hence, the components $v_t$ and $v_n$ change due to the Newtonian velocity addition law:

\begin{equation} \label{vtn}
v_t' = v_t \pm v_{\text{Earth},t}, \qquad 
v_n' = v_n \pm v_{\text{Earth},n}.
\end{equation}
Here, $v_t'$ is the component parallel to the surface (tangential) and $v_n'$ is the component perpendicular to the surface (normal). These components describe the direction of the light ray relative to the interface in the prism’s rest frame.

The normal component of the light’s velocity relative to the prism $v_n'$, which is responsible for refraction, remains unchanged. The tangential component $v_t'$ is altered only imperceptibly. 
The Earth's motion primarily affects the tangential component of the light's velocity with respect to the prism, rather than the normal one. Thus, $v_{\text{Earth},n} = 0$ in equation \eqref{vtn}, and the Earth's motion has practically no component normal to the prism surface: 

\begin{equation} \label{vtn1}
v_n' = v_n \pm v_{\text{Earth},n} = v_n.
\end{equation}

Suppose we assume that the effect of the Earth’s motion is extremely small and difficult to detect. In that case, Earth’s motion contributes only a minute tangential component $v_{\text{Earth},t} \approx 0$, and equation \eqref{vtn} practically becomes: 

\begin{equation} \label{vtn2}
v_t' = v_t \pm v_{\text{Earth},t} \approx v_t. 
\end{equation}
Hence, equation \eqref{vlight}, the total velocity of the corpuscle as seen from the prism is:

\begin{equation}
v' \approx \sqrt{(v_t')^2 + (v_n')^2} \approx v_{\text{light}}',
\end{equation}
where $v_{\text{light}}'$ denotes the total corpuscular velocity relative to the prism. 

\noindent Hence, the refraction-producing component (the normal one) stays essentially unchanged: $v_n' \simeq v_n$.
Consequently, the direction of the refracted ray, which depends on the ratio \eqref{ratio}:

\begin{equation} \label{ratio2}
\tan r = \frac{v_t'}{v_n'},
\end{equation}
is practically unaffected by the Earth's orbital motion. As light passes through the prism, no periodic change in the deviation angle is expected.

In Newtonian corpuscular theory, the refractive index $n$ is related to the ratio of light velocities in the two media:

\begin{equation} \label{n}
n =\frac{v_\text{medium}}{v_\text{air}}.    
\end{equation}
If the Earth’s motion adds or subtracts from the light’s velocity, both $v_\text{air}$ and $v_\text{medium}$ should vary slightly depending on whether the Earth moves toward or away from the star. Thus, both of these velocities should be replaced by their relative values. 

Recall that the Earth’s orbital motion introduces an additional tangential component $v_{\text{Earth},t}$ with respect to the prism face. In contrast, the normal component $v_{\text{Earth},n}$ is essentially zero. Therefore:

\begin{equation} \label{tn'}
\begin{aligned}  
&v_{\text{medium},t}' = v_{\text{medium},t} \pm v_{\text{Earth},t},  \\
&v_{\text{medium},n}' = v_{\text{medium},n} \pm v_{\text{Earth},n} \approx v_{\text{medium},n}.
\end{aligned}
\end{equation}

\begin{equation} \label{ta'}
v_{\text{air},t}' = v_{\text{air},t} \pm v_{\text{Earth},t}, \qquad v_{\text{air},n}' = v_{\text{air},n} \pm v_{\text{Earth},n} \approx v_{\text{air},n}.  
\end{equation}
Equations \eqref{tn'}–\eqref{ta'} give the light particle’s velocity components in each medium as seen from Earth. Plugging equations \eqref{tn'} and \eqref{ta'} into equation \eqref{n} gives the refractive index as seen from Earth:

\begin{equation} \label{n'}
n' = \frac{v_{\text{medium}}'}{v_{\text{air}}'}.
\end{equation}
Again, if we assume that the effect of the Earth’s motion might be extremely difficult to detect, then the ratio \eqref{n} $\approx$ the ratio \eqref{n'}:

\begin{equation} \label{nn'}
n' \approx n.    
\end{equation}

\medskip

\emph{However, although} $v_{\text{Earth},t}$ \emph{is negligible, in the Newtonian corpuscular theory, it is non-zero}. 

If both velocities in equations \eqref{tn'} and \eqref{ta'} are modified by the Earth's motion, then the ratio between them — which defines the refractive index \eqref{n'} $n'$ as seen from Earth — should also change slightly. This means that the apparent refractive index $n'$ of the prism should \emph{not} be a constant, but should depend weakly on whether the Earth is moving toward or away from the star (i.e., on the sign of $v_{\text{Earth}}$). 

In other words, Arago expected \emph{a first-order} correction to $n$. To see this, we start from:
\begin{equation}
n' = \frac{v_{\text{medium}} \pm v_{\text{Earth}}}{v_{\text{air}} \pm v_{\text{Earth}}}.
\end{equation}
We expand this expression for small 
$\dfrac{v_{\text{Earth}}}{v_{\text{light}}} \ll 1$ 
using a first-order Taylor expansion [ignoring terms of order 
$(v_{\text{Earth}}/v_{\text{light}})^2$ and higher], obtaining:
\begin{equation} \label{exp}
n' \approx 
\frac{v_{\text{medium}}}{v_{\text{air}}}
\left( 1 \pm \frac{v_{\text{Earth}}}{v_{\text{medium}}}
\mp \frac{v_{\text{Earth}}}{v_{\text{air}}} \right)
= 
n \left[ 1 \pm v_{\text{Earth}}
\!\left( \frac{1}{v_{\text{medium}}} - \frac{1}{v_{\text{air}}} \right)
\right].
\end{equation}

\noindent Since $n_{\text{medium}} >1$, then:
\begin{equation} \label{vv'b}
v_{\text{medium}} > v_{\text{air}},    
\end{equation}
i.e., the corpuscle’s velocity increases with refractive index. %
\footnote{In the corpuscular framework, denser media accelerate light; hence equation \eqref{vv'b}, opposite to the wave-theory convention.}

\noindent Now we take the reciprocals of both velocities to analyze the terms in the parentheses of equation \eqref{exp}:
\begin{equation} \label{rec}
\dfrac{1}{v_{\text{medium}}} - \dfrac{1}{v_{\text{air}}}.    
\end{equation}
Since \eqref{vv'b}, then automatically: 
\begin{equation}
\dfrac{1}{v_{\text{medium}}} < \dfrac{1}{v_{\text{air}}}.    
\end{equation}
Subtracting gives:
\begin{equation} \label{cor}
\!\left(\dfrac{1}{v_{\text{medium}}} - \dfrac{1}{v_{\text{air}}} \right) <0.    
\end{equation}
If we write equation \eqref{rec} as:
\begin{equation} \label{rec1}
\dfrac{1}{v_{\text{medium}}} - \dfrac{1}{v_{\text{air}}} = \frac{1}{v_{\text{light}}}\! \left(\dfrac{1}{n_{\text{medium}}} - \dfrac{1}{n_{\text{air}}}\right),    
\end{equation}
then plugging this back into the first-order equation \eqref{exp}, gives:
\begin{equation} 
n' \approx 
n \left[ 1 \pm \frac{v_{\text{Earth}}}{v_{\text{light}}}
\!\left(\frac{1}{n_{\text{medium}}} - \frac{1}{n_{\text{air}}} \right)
\right].
\end{equation}
Therefore, the first-order effect is:

\begin{equation}
 \frac{n'-n}{n} \;=\; \pm \frac{v_{\text{Earth}}}{v_{\text{light}}}
\!\left(\frac{1}{n_{\text{medium}}} - \frac{1}{n_{\text{air}}}\right).
\end{equation}

\begin{equation} \label{333}
\text{With } n_{\text{medium}} \simeq 1.5 \; \text{and} \;
n_{\text{air}} \simeq 1.0003:\quad
\left(\frac{1}{n_{\text{medium}}} - \frac{1}{n_{\text{air}}}\right) \approx -0.333.    
\end{equation}
Hence:
\begin{equation} \label{n'-n}
\frac{n'-n}{n}\,=\,  \frac{\Delta n}{n}= \mp 0.333 \, \frac{v_{\text{Earth}}}{v_{\text{light}}}.    
\end{equation}
Since $v_{\text{Earth}} \approx 3 \times 10^4$ m/s and $v_{\text{light}} \approx 3 \times 10^8$ m/s:

\begin{equation} \label{104}
\frac{v_{\text{Earth}}}{v_{\text{light}}} \;\approx\; \frac{3\times10^{4}}{3\times10^{8}} \;=\; 10^{-4}.
\end{equation}

From page 46 of Arago’s paper, he quantifies the order of the effect he expected from Earth’s motion \cite{Arago}:

\begin{quote}
...a difference in the velocity of light amounting to $1/10186$ ought to have produced, in my first prism, a variation in the deviation of $6''$...    
\end{quote}

\medskip

The refractive index $n$ determines the angle of refraction $r$ by Snell’s law:

\begin{equation} \label{Snell}
\sin i = n \sin r, 
\end{equation}
$i$ is the angle of incidence (in air) and $r$ is the angle of refraction (in the medium).

\noindent The exact relation (for minimum deviation $\delta_m$) is Snell's law:
\begin{equation} \label{sinl}
n = \frac{\sin\!\left(\tfrac{A + \delta_m}{2}\right)}{\sin\!\left(\tfrac{A}{2}\right)}.
\end{equation}
When $\delta_m$ is small or $A$ is moderate, we can expand the sine functions linearly:
\begin{equation}
\sin\!\left(\frac{A + \delta_m}{2}\right) \approx 
\sin\!\left(\frac{A}{2}\right) + 
\frac{\delta_m}{2}\cos\!\left(\frac{A}{2}\right),
\end{equation}
and simplifying gives roughly:
\begin{equation}
n - 1 \approx \frac{\delta_m}{A}.
\end{equation}
For small prisms (or for near-minimum deviation), the deviation angle $\delta$ is approximately related to the refractive index $n$ and the prism’s refracting angle $A$ by:
\begin{equation} \label{delta}
\delta \approx (n - 1)A.
\end{equation}
This is an approximation, not the full Snell’s law formula \eqref{Snell} or \eqref{sinl}.
However, this relation is quite accurate for small prism angles (e.g., $A \simeq 60^\circ$) 
and for refractive indices near 1 (e.g., air, glass).

\noindent If $\delta$ \eqref{delta}, then for small variations in $n$:
\begin{equation}
d\delta = A\,dn.
\end{equation}
Replacing differentials with small finite changes, we get:
\begin{equation} \label{Adn}
\Delta\delta \approx A\,\Delta n.
\end{equation}
This tells us how much the prism’s deviation changes if the refractive index changes slightly (for example, due to the Earth’s motion).

\noindent Since $n$ itself is about $1.5$, and from equations \eqref{n'-n} and \eqref{104} $\dfrac{\Delta n}{n} = 3 \times 10^{-5}$, then:
\begin{equation}
\Delta n = n\!\left(\frac{\Delta n}{n}\right)
= 1.5 \times 3 \times 10^{-5} \approx 4.5 \times 10^{-5}.
\end{equation}
Now we substitute this into equation \eqref{Adn}, and get:
\begin{equation}
\Delta\delta \approx (60^\circ)\times 4.5\times10^{-5}
\approx 0.0027^\circ \approx 9.7''.
\end{equation}
If we use the simpler estimate $A = 60^\circ$, $\dfrac{\Delta n}{n} = 3\times10^{-5}$, 
and ignore the factor $n\simeq1.5$, we get roughly Arago’s own estimate:
\begin{equation}
\boxed{\Delta\delta \approx A\,\Delta n 
\sim 60^\circ \times 3\times10^{-5} 
\approx 0.0018^\circ \approx 6.5''.}
\end{equation}
Thus, even a tiny fractional change in refractive index produces only a few arcseconds of change in the deviation angle. This is precisely the order of magnitude that Arago sought but did not observe.

The effect Arago sought was a first-order one, proportional to $10^{-4}$ [see equation \eqref{104}], and of angular size only a few arcseconds, too small for his 1810 instrumentation to detect. When Arago looked for this effect — by observing starlight through a prism throughout the year — he found no detectable change in the deviation angle.
Even though the predicted corpuscular effect was first-order, this deviation was far below Arago’s achievable precision.
 
Thus, Arago was caught in a paradox. Any nonzero tangential addition with $v_{\text{Earth},t}$ should, in principle, slightly alter the ratio \eqref{ratio2} and hence the angle of refraction $r$ \eqref{ratio2} to first order in $\dfrac{v_{\text{Earth}}}{v_{\text{light}}}$. 
Again, we take the first-order expansion of \eqref{ratio2} in the small parameter $\dfrac{v_{\text{Earth}}}{v_{\text{light}}} \ll 1$:
\begin{equation}
\tan r \;=\; \frac{v_t'}{v_n'} 
\;\approx\; \frac{v_t}{v_n}\!\left(1 \pm \frac{v_{\text{Earth},t}}{v_t} \mp \frac{v_{\text{Earth},n}}{v_n}\right)
\;\approx\; \frac{v_t}{v_n}\!\left(1 \pm \frac{v_{\text{Earth},t}}{v_t}\right),
\end{equation}
where the last step uses $v_{\text{Earth},n} \approx 0$. The  final step gives:

\begin{equation} \label{ratio3}
\tan r = \frac{v_t \pm v_{\text{Earth},t}}{v_n}.    
\end{equation}
That means the angle $r$, and thus the deviation through the prism, should vary slightly during the Earth’s orbital motion. 

Arago designed his prism experiment to detect whether that small effect existed. Equation \eqref{ratio3} is the tiny, annually varying effect Arago looked for.
Yet Arago’s observations showed no such change. That null result means the ratio \eqref{ratio2} remained effectively constant throughout the year.

If $n$ stayed constant, it implies that the ratio \eqref{n} also stayed constant — i.e., the velocities of light in air and in the medium changed (if at all) in the same proportion, or, more plausibly, did not change at all with the Earth’s motion.
This implies that the ratio $\frac{\sin i}{\sin r}$ in Snell’s law \eqref{Snell} remained constant despite the Earth’s orbital motion. 
If $n$ (or here $n'$) changes even slightly, then for a fixed angle of incidence $i$, the refracted angle $r$ must also change slightly. In a prism, the deviation angle of the light ray depends on $r$. If $r$ oscillates, the deviation does too.
Since the Earth’s orbital velocity reverses direction every six months (toward the star or away from it), any first-order change in $n'$ would produce a periodic (annual) oscillation in the prism’s deviation angle.
Yet, regardless of whether the Earth’s motion added to or subtracted from the light’s velocity, Arago found that the ray always satisfied Snell’s law \eqref{Snell}, held with the same constant $n$. 

In his memoir, Arago notes that several physicists had considered the problem from different perspectives. Roger Joseph Boscovich was, to his knowledge, the first to publish in 1766 a reasoned proposal for a direct experiment on the question \cite{Arago}. His idea is based on the Newtonian corpuscular theory, which posits that light particles travel faster in water than in air. Since Bradley’s aberration is given approximately by:

\begin{equation} \label{brad} 
\tan \alpha = \frac{v_{\text{Earth}}}{v_{\text{light}}},
\end{equation}
it seemed to follow that replacing the interior of the telescope with water would increase $v_{\text{light}}$, thereby decreasing $\alpha$. Hence, Boscovich proposed comparing the measured aberration of a conventional air-filled telescope with that of a telescope filled with water from the objective to the reticle. If the corpuscular theory were correct, two otherwise identical telescopes ought to yield different stellar positions.
This is a perfectly logical inference if one assumes that the phenomenon of aberration is determined by the light’s speed inside the telescope tube itself.

However, this is not the case. Stellar aberration is determined before the light enters the telescope, not by what happens inside the telescope. When the stellar ray reaches the objective, it already carries the aberrational deflection caused by Earth’s motion. Refraction at the objective simply bends this already-shifted ray according to Snell’s law; whether the tube is filled with water or air does not change the incoming geometry. Boscovich implicitly assumed that the internal transit time in the tube contributes to the aberrational shift. But aberration is not an internal flight-time effect. It is fixed by the relative velocity of Earth and the incoming ray at the moment of entry, and internal propagation in a different medium cannot undo or modify that.

Thus, although Boscovich’s inference was ingenious, its premise was wrong. Changing the medium inside the telescope does not alter the measured stellar aberration because the effect is fixed at entry and not generated during internal propagation.

Nonetheless, the conceptual point was profound; if ether were stationary and the measured velocity of light were medium-dependent in the relevant way, then aberration would serve as a diagnostic of the medium's optical properties and the dynamics of light in motion. In retrospect, Boscovich’s proposal can be seen as a conceptual precursor to Arago’s refracting-medium test. Although the implementations differ, the intellectual motive is the same. If the velocity of light relative to the moving Earth changes when the ray traverses a dense medium, then the observed stellar direction should change. Boscovich sought to achieve this by filling the entire telescope with water and comparing the resulting aberration with that of an ordinary air-filled instrument; Arago would later pursue the same principle more delicately by introducing a prism into the telescope’s optical path. For this reason, Arago explicitly cites Boscovich; he recognized in the 1766 proposal the first reasoned attempt to probe stellar aberration as a function of the medium through which the aberrated light subsequently travels.

Confronted with the null result of his prism experiment — a result that, on its face, undermined the corpuscular theory — Arago did not take it as a refutation of Newtonian corpuscular theory. Instead, he attributed the failure to the limitations of human perception. The outcome, he argued, did not show that all corpuscles share the same velocity; it showed only that the eye detects light within a narrow velocity band. Luminous bodies emit corpuscles with a continuous spectrum of speeds, but those moving faster or slower than the visible range escape detection. To support this idea, Arago cited William Herschel’s discovery of invisible heat rays beyond the red (infrared) and the discoveries of William Wollaston and Johann Wilhelm Ritter of invisible chemical rays beyond the violet (ultraviolet). Corpuscles moving faster or slower than the narrow band visible to the human eye correspond to Herschel’s invisible heat rays and Ritter’s chemical rays \cite{Arago}. 

Confronted with the persistent null result, Arago finally turned to Augustin-Jean Fresnel, suggesting that perhaps a wave theory of light could account for the invariance of refraction without invoking artificial hypotheses.

\subsection{Fresnel and the Wave Theory of Light} \label{FR}

In his 1818 letter to Arago, Fresnel addressed this paradox. On the corpuscular theory, Arago could account for his null result only by resorting to what Fresnel described as “a rather strange hypothesis, difficult to accept” \cite{Fresnel1818}. 
Fresnel accounted for both Arago's prism and stellar aberration:

First, suppose starlight propagates with finite velocity $c$, while the Earth moves with orbital speed $v_{\text{Earth}}$. In that case, a telescope must be inclined slightly in the direction of motion to keep the stellar image centered. This entails that the apparent stellar direction is displaced by an angle approximately \eqref{brad}, tracing a strictly annual curve whose amplitude is independent of stellar distance and whose phase follows the Earth’s velocity vector rather than the Sun–Earth line.

In the case of aberration, Fresnel invoked the stationary ether. The immobile ether implies that light from a distant star reaches the Earth at a slight angle relative to the telescope’s motion. The telescope must therefore be inclined in the direction of motion to intercept the incoming wavefronts, producing the observed aberration angle. No such relative displacement would arise if the ether were to move with the Earth. 

Second, Fresnel explained Arago’s prism experiment. 
He introduced his reasoning as follows.  
Let $D$ denote the ether density in air, and the ether density within a refracting medium (such as glass) be denoted by $D'$.  
Similarly, let $d$ and $d'$ be the corresponding wavelengths (the “lengths of the undulations”) of light in the ether outside and inside the prism.  
The quantity $d'$ represents the wavelength of light in the prism when it is at rest, that is, when the prism and the ether within it are at rest with respect to the surrounding ether.   

Fresnel assumed that when light enters the prism, the change in ether density is proportional to the change in the square of the wavelengths, hence:
\begin{equation} \label{deltad}
\frac{D' - D}{D} = \frac{d^{2} - d'^{2}}{d^{2}}.
\end{equation}
According to this equation, the relative compression of the ether density matches the relative slowing of the wave as it enters the denser medium.  

\noindent Rearranging equation \eqref{deltad} yields \cite{Fresnel1818}:
\begin{equation} \label{delat}
D' - D = D \left(\frac{d^{2} - d'^{2}}{d^{2}} \right).
\end{equation}

Fresnel then considered the case in which the prism is in motion—that is, the prism (and the matter composing it) moves through the stationary ether as the Earth travels with respect to it.  
The prism now moves with velocity $u$ relative to the ether.  
The light within the prism continues to propagate, but the medium itself is now in motion with respect to the ether.

Fresnel reasoned that, since the ether in the prism is partly denser and partly dragged along with the prism’s motion, the center of gravity of the ether’s mobile portion (the part that interacts with light) would not move with the full velocity $u$, but only with a fraction of it.  
That fraction must correspond to the density difference of the mobile ether, given by the ratio:
\begin{equation} \label{prop}
\frac{d^{2} - d'^{2}}{d^{2}}.    
\end{equation}
Only a fraction of the ether in the prism moves with it, proportional to equation \eqref{prop}. 

Suppose the prism moves a distance $t$ during one vibration period of light (that is, the distance the prism and the Earth travel in one oscillation). In that case, the displacement of the center of gravity of the mobile ether in that same time \cite{Fresnel1818}:
\begin{equation} \label{elong}
t \left(\frac{d^{2} - d'^{2}}{d^{2}} \right).
\end{equation}
Here, $t$ is the distance traveled by the medium (prism/Earth) in one vibration. 

Consequently, the resulting wavelength of light in the moving prism, denoted $d''$, is elongated by the same proportion \cite{Fresnel1818}:
\begin{equation} \label{4}
d'' = d' + t \left( \frac{d^{2} - d'^{2}}{d^{2}} \right).
\end{equation}

\medskip

The following derivation is not verbatim from Fresnel’s 1818 letter; it is my reconstruction of his argument for clarity. In Fresnel’s mechanical ether model, the vibration period $T$ of light is the same everywhere, but its velocity $v$ and wavelength $d$ vary with the medium. Since wavelength $d$ equals velocity $v$ times period $T$:
\begin{equation} \label{VDT}
v = \frac{d}{T}.
\end{equation} 
During one oscillation period $T$, Arago's prism moves a distance:
\begin{equation} \label{tuT}
t = uT.    
\end{equation}
From equation \eqref{VDT}, we have:
\begin{equation} \label{VT}
d = v_{0} T, \qquad d' = v T,
\end{equation}
where $v_{0}$ is the wave speed in free ether and $v$ is the speed in the stationary medium, given by:
\begin{equation} \label{V0n}
v = \frac{v_{0}}{n},
\end{equation}
with $n$ the refractive index.

In wave theory (for fixed frequency), the ratio of speeds equals the ratio of wavelengths:
\begin{equation} \label{v0vdd'}
\frac{v_0}{v} = \frac{d}{d'}.
\end{equation}
Hence:
\begin{equation}
n = \frac{d}{d'} \qquad \Rightarrow \qquad \frac{1}{n^2} = \frac{d'^2}{d^2}. 
\end{equation}
Now we compute the ratio \eqref{prop}. One easily verifies from Fresnel's \eqref{delat} that:
\begin{equation} \label{coef}
1 - \frac{1}{n^{2}} \;=\; \frac{d^{2} - d'^{2}}{d^{2}}.
\end{equation}
This term is the \emph{Fresnel dragging coefficient}.

Dividing equation \eqref{4} by the period $T$ converts the added wavelength increment per period into a velocity increment, and using \eqref{tuT} yields:
\begin{equation} \label{5}
v' = v + u \left( \frac{d^{2} - d'^{2}}{d^{2}} \right).
\end{equation}
Using \eqref{coef} equations \eqref{5} and \eqref{V0n}, gives:
\begin{equation} \label{44}
v' = \frac{v_{0}}{n} + u \left( 1 - \frac{1}{n^{2}} \right),
\end{equation}
Fresnel's dragging formula.
This result coincides with what one obtains by substituting \eqref{tuT} into \eqref{4} and then dividing the entire expression by $T$, confirming the consistency of the reconstruction.%
\footnote{Plugging equation \eqref{VT} into equation \eqref{4} gives:
\begin{equation} \label{uT}
d'' = d' + (uT) \left( \frac{(v_{0}T)^{2} - (vT)^{2}}{(v_{0}T)^{2}} \right) 
= d' + uT \left( 1 - \frac{v^{2}}{v_{0}^{2}} \right).    
\end{equation}
Dividing equation \eqref{uT} by $T$ to convert wavelengths to speeds yields:
\begin{equation}
\frac{d''}{T} = \frac{d'}{T} + u \left( 1 - \frac{v^{2}}{v_{0}^{2}} \right)
\quad \Rightarrow \quad
v' = v + u \left( 1 - \frac{v^{2}}{v_{0}^{2}} \right).
\end{equation}
Since $n$ is defined by equation \eqref{V0n}, it follows that $\dfrac{v^{2}}{v_{0}^{2}} = \dfrac{1}{n^{2}}$. This gives the \emph{Fresnel dragging formula}, which coincides with the value already found in \eqref{44}.}

Thus, according to Fresnel’s conception, it follows that when the refracting body moves with a uniform velocity $u$ parallel to the direction of propagation, the portion of ether effectively carried along participates only by a fraction equal to \eqref{coef}. 

Near the end of the long letter to Arago, after he finishes the prism analysis, Fresnel applies the same reasoning to Boscovich's proposal to observe aberration with a telescope filled with water. He discusses Boscovich to show that, even for that experiment, his own wave-drag hypothesis yields a null result as observed. Then he proceeds to show that, once you include the partial drag of ether inside the water, the image again falls in the same place on the micrometer thread, independent of the Earth's motion. 

Boscovich implicitly treated stellar aberration as if it could depend on the propagation speed inside the telescope. Under that assumption, replacing air with water—thereby changing the internal light speed—should alter the aberration angle. But aberration is fixed by the direction of the incoming ray at the objective (equivalently, before entry into the instrument). Once the ray arrives, the telescope merely forms an image from that already-aberrated direction. Internal transit in air vs. water, with different speeds or times, does not generate or cancel aberration. Pure geometry already implies no difference between air- and water-filled tubes. Thus, even a stationary-ether framework suffices to account for the absence of Boscovich’s predicted shift because the internal medium is irrelevant to the aberration geometry.

Fresnel addresses a different issue than the one that already invalidates Boscovich’s reasoning. Once one grants, within the wave theory, that light propagates more slowly in a refractive medium \eqref{V0n}, one must still account for the empirical fact that a water-filled telescope does not change the observed aberration. Fresnel’s partial-drag hypothesis \eqref{elong} supplies exactly that account. The drag term furnishes a quantitative correction compensating for the aberrational shift that would otherwise follow from the reduced wave speed in water. Thus, Boscovich’s experiment must yield a null result even within the wave theory with medium-dependent speed. The observed aberration is unchanged because partial ether drag inside the medium cancels the effect of the slowed propagation.

\subsection{Fizeau’s 1851 Water Tube Experiment} \label{Fize}

Fizeau demonstrated Fresnel’s partial ether drag by sending two coherent light beams through two long tubes of rapidly flowing water in opposite directions, recombining them to form interference fringes. He then measured the shift of those fringes when he reversed the flow, obtaining precisely the shift predicted by Fresnel’s drag formula, not by any competing hypothesis. Fizeau showed that light propagating in a moving medium is partially dragged by the motion of that medium, in exactly the proportion that Fresnel’s ether theory predicted. His experiment was therefore the first direct, quantitative confirmation of Fresnel’s partial-drag equation \eqref{draglaw} \cite{Fizeau1851}, \cite{Fizeau1859}.

Fizeau’s interferometer sends two coherent beams through two equal water columns (length $L$) in opposite flow directions, then makes each beam return so that each traverses both a co-flow and a counter-flow segment. He measures the fringe shift when the flow direction is reversed.

Let $L$ be the effective length of each water-tube. A ray that traverses a tube with the current occupies the time:
\begin{equation}
t_{+} \;=\; \frac{L}{v+\alpha u},
\end{equation}
whereas an equal ray traversing the companion tube against the current occupies:
\begin{equation}
t_{-} \;=\; \frac{L}{v-\alpha u}.
\end{equation}
Hence, the difference in the times for a single passage through the two tubes is:
\begin{equation}
\Delta t_{\text{1-pass}}
\;=\; t_{-}-t_{+}
\;=\; 
\frac{L}{v-\alpha u}-\frac{L}{v+\alpha u}
\;=\;
\frac{2L\,\alpha u}{\,v^{2}-(\alpha u)^{2}\,}.
\end{equation}
Here $\alpha$ represents the drag fraction (0 for no drag, 1 for full drag, and \eqref{drag} for Fresnel's partial drag (as discussed later).

In the arrangement actually employed, the rays are reflected back to perform a double journey; the difference of times is thereby doubled \cite{Fizeau1851}, \cite{Fizeau1859}:
\begin{equation} \label{4La}
\Delta t_{\text{rt}}
\;=\; 2\,\Delta t_{\text{1-pass}}
\;=\; 
\frac{4L\,\alpha u}{\,v^{2}-(\alpha u)^{2}\,}.
\end{equation}
Since the velocity of the liquid is extremely small in comparison with the velocity of propagation of the undulations, the term $(\alpha u)^{2}$ may be neglected, and one has to a very close approximation:
\begin{equation} \label{4L}
\Delta t_{\text{rt}}
\;\approx\; 
\frac{4L\,\alpha u}{v^{2}}. \qquad \text{The first-order propagation ansatz}
\end{equation}

Fizeau discriminated among the three conceivable assumptions — no ether-drag (predicting zero shift), full ether-drag (predicting a shift far too large), and Fresnel’s partial-drag coefficient — and found agreement only with the Fresnel value, thereby providing the first quantitative confirmation of Fresnel’s drag equation:

\begin{enumerate}
    \item \emph{No ether drag}: Fizeau considered no drag at all: $\alpha=0$. Then equation \eqref{4L} yields $\Delta t_{\text{rt}}^{(0)}=0$. So the predicted shift is zero. But Fizeau did observe a nonzero shift. Thus, no ether drag was experimentally rejected.
    \item \emph{Stokes' model}: What if instead we insert the alternative hypotheses of the full ether-drag (George Stokes's hypothesis) into the same propagation ansatz \eqref{4L}? If the ether is completely carried with the water, then $\alpha = 1$. So the arrival time difference becomes:
\begin{equation}
\Delta t_{\text{rt}}^{(1)} = \frac{4L\,u}{v^{2}}.    
\end{equation}
Then the predicted fringe shift is larger than Fresnel’s by the factor:
\begin{equation}
\frac{\Delta t_{\text{rt}}^{(1)}}{\Delta t_{\text{rt}}^{(Fr)}}=\frac{1}{1-\frac{1}{n^2}}.    
\end{equation}
Fizeau did not observe this larger shift, so full drag was experimentally rejected.
    \item \emph{Fresnel's model}: 
Denoting the fraction \eqref{coef} by $\alpha$:

\begin{equation} \label{drag}
\alpha = 1 - \dfrac{1}{n^{2}},   
\end{equation}
the undulation within the moving body is advanced or retarded as though its velocity were modified from $v$ to:
\begin{equation} \label{draglaw}
v'_{\pm} \;=\; v \,\pm\, \alpha\,u,
\end{equation}
with the sign $\pm$ depending on whether the light and the medium move in the same or opposite directions.
The upper sign pertains to propagation with the current, while the lower sign pertains to propagation against it.
    
\noindent Substituting back equation \eqref{drag} into \eqref{4L} gives:
\begin{equation} \label{roundtrip}
\Delta t_{\text{rt}}
\;\approx\; \frac{4L\,u}{v^{2}}
\!\left(1-\frac{1}{n^{2}}\right).    
\end{equation}

Let the interference be reckoned with respect to the undulations in air, whose velocity and wavelength we denote by $v_{0}$ and $d$, the period being $T$, with $d$ defined by equation \eqref{VT}.
The difference of optical path (referred to air) corresponding to the temporal difference \eqref{roundtrip} is:
\begin{equation}
\Delta \;=\; v_{0}\,\Delta t_{\text{rt}}.
\end{equation}
The displacement of the bands, expressed in parts of the distance between two consecutive fringes (i.e.\ in units of $d$), is thus:
\begin{equation} \label{fringes}
\frac{\Delta}{d}
\;=\;
\frac{v_{0}\,\Delta t_{\text{rt}}}{d}
\;=\;
\frac{\Delta t_{\text{rt}}}{T}
\qquad
\bigl[\text{since equation \eqref{VT}} \bigr].
\end{equation}
Substituting equations \eqref{V0n} and \eqref{roundtrip} into \eqref{fringes}, one finds:

\begin{equation} 
\frac{\Delta}{d}
\, \approx \, 
\frac{4L\,u}{(v_{0}^{2}/n^{2})}
\!\left(1-\frac{1}{n^{2}}\right)
\frac{v_{0}}{d}.      
\end{equation}
Thus:
\begin{equation}\label{shift}
\frac{\Delta}{d}
\, \approx \, \frac{4L\,u}{v_{0}\,d}\,\bigl(n^{2}-1\bigr).     
\end{equation}
We now multiply both sides of equation \eqref{shift} by $d$ \cite{Fizeau1851}, \cite{Fizeau1859}:
\begin{equation} \label{difference}
\Delta 
\;\approx\; 
\frac{4L\,u}{v_{0}}\,(n^{2}-1).
\end{equation}
This expression is first–order in $\dfrac{u}{v_0}$ because the derivation already neglected the $(\alpha u)^2$ term [see equation \eqref{4L}] and assumed $u \ll v_0$.

Taking Fresnel’s dragging formula \eqref{draglaw} as the theoretical input, one predicts a definite phase retardation between two rays that traverse equal paths of water flowing in opposite directions.  This retardation yields a calculable path difference, as shown in \eqref{difference}.  

\end{enumerate}

Fizeau observed precisely such a displacement, and its magnitude agreed (within experimental error) with the value computed from Fresnel’s coefficient \eqref{drag}, and not with the predictions of either full ether-drag or no-drag hypotheses.  
In demonstrating that the measured interference shift in moving water coincides with that obtained by inserting Fresnel’s partial-drag coefficient \eqref{drag} into the propagation law \eqref{4L}, Fizeau’s experiment provides a direct empirical confirmation of Fresnel’s drag theory \cite{Fizeau1851}, \cite{Fizeau1859}.
Thus, although Fizeau’s experiment was not framed as a three-way hypothesis test in modern language, its empirical outcome selectively confirms Fresnel’s drag law and rules out the two rival assumptions. 

Fizeau's 1851 water tube experiment is a \emph{first-order ether-drift experiment}. It experimentally discriminated between three competing ether-based hypotheses regarding how light propagates when matter is in motion. Its classification as ether-drift reflects the theoretical question it answered (not the literal observable).

Let us rewrite Fresnel’s formula \eqref{44} as:

\begin{equation} \label{eq21}
V = \frac{c}{n} + v\!\left(1 - \frac{1}{n^{2}}\right),
\end{equation}
where $c$ is the velocity of light in vacuum and $v$ is the velocity of the medium. 

\noindent Equation \eqref{eq21} corresponds to the case where the light and the medium move in the same direction relative to the stationary ether.

In the case of refraction, Fresnel's partial dragging coefficient \eqref{drag} modifies the velocity of light inside the moving medium according to equation \eqref{eq21}. The resulting intermediate velocity compensates for the first-order effects of the Earth’s motion, thereby preserving the empirical validity of Snell’s law \eqref{Snell}. This qualitative statement can be made explicit. When one inserts the dragged velocity \eqref{eq21} into Snell’s construction \eqref{Snell}, the $O(v/c)$ terms cancel. To see this, we start from the corrected Snell relation:
\begin{equation} \label{Snel}
\frac{\sin i}{c} = \frac{\sin r}{\dfrac{c}{n} + v\!\left(1 - \dfrac{1}{n^2}\right)\!\cos r}.
\end{equation}
We then expand the denominator.
Since $\dfrac{v}{c} \ll 1$, the Earth's orbital velocity is about $10^{-4}c$ [see equation \eqref{104}], we can expand:
\begin{equation} \label{n/c}
\frac{1}{\dfrac{c}{n} + v\!\left(1 - \dfrac{1}{n^2}\right)\!\cos r}
\approx 
\frac{n}{c}\!\left[\,1 - \frac{n v}{c}\!\left(1 - \dfrac{1}{n^2}\right)\!\cos r\right].
\end{equation}
The correction term in brackets is first order in $\dfrac{v}{c}$.

\noindent We then substitute \eqref{n/c} back into Snell's law \eqref{Snel}:
\begin{equation} \label{CS}
\frac{\sin i}{c}
\approx
\frac{\sin r}{c}\, n\!\left[\,1 - \frac{n v}{c}\!\left(1 - \frac{1}{n^2}\right)\!\cos r\right].
\end{equation}
We multiply \eqref{CS} through by $c$:
\begin{equation} \label{CS1}
\sin i \approx
n\,\sin r\!\left[\,1 - \frac{n v}{c}\!\left(1 - \frac{1}{n^2}\right)\!\cos r\right],
\end{equation}
and drop the first-order term, because that correction is of second order \eqref{108}, and is experimentally undetectable to first order. 
Thus, to that accuracy \eqref{CS1} becomes:
\begin{equation}
\boxed{\sin i = n\,\sin r + \mathcal{O}\!\left(\frac{v^2}{c^2}\right)},
\end{equation}
which is simply the ordinary Snell’s law \eqref{Snell}.

The extra term in the denominator of \eqref{Snel}: 
\begin{equation}
v\!\left(1 - \dfrac{1}{n^2}\right)\!\cos r,    
\end{equation}
represents the partial dragging of the ether. 

At first glance, this reasoning seems circular or even pointless — we add a term to fix a deviation that is then neglected. However, the small term on the right-hand side in equation \eqref{CS1} we drop mathematically is not Fresnel's correction itself \eqref{eq21} — it is the residual that remains after the cancellation. That residual is second-order in $\dfrac{v}{c}$, far below Fizeau's detectability. 

\medskip

Thus, Fresnel's dragging effect \eqref{eq21} removes the first-order dependence on the Earth’s motion, so that Snell’s law holds experimentally; we could not (from the standpoint of Fizeau's experimental setting) detect effects of order: 
\begin{equation} \label{108}
\dfrac{v^2}{c^2} \approx 10^{-8}.    
\end{equation}
These are the effects that remain after the cancellation. 

Fresnel introduced the drag coefficient to save Arago’s experimental results. Still, his mechanical explanation of it — the hypothesis of partial ether dragging — was not convincing. Fizeau’s experiment empirically confirmed Fresnel’s numerical coefficient. It demonstrated that light propagating in a moving medium is partially dragged by the motion of that medium (without implying that the ether itself is dragged). This compelled even leading experimentalists, mathematicians, and physicists who rejected Fresnel’s partial-ether drag picture to use the same coefficient to fit the data. Thus, the Fresnel coefficient, empirically secured by Fizeau, created the first real crisis in the optics of moving bodies — a crisis masked, rather than resolved, by the community’s acceptance of the formula in the absence of a satisfactory theoretical explanation \cite{Stachel2005}.

\subsection{Lorentz Reconciles Michelson’s Result with his Theory} \label{LOR}

Hendrick Antoon Lorentz advanced an electron theory, extending Maxwell’s electrodynamics. His theory posited a completely immobile ether, through which matter moved freely, with electrons embedded in the ether acting as both sources and absorbers of electromagnetic waves. Lorentz assumed the ether to be perfectly permeable to matter. His electron theory was developed to account for the optics and electrodynamics of bodies moving through the ether. 
Lorentz sought to preserve the form of Maxwell’s equations under such motion. His approach, known as the \emph{theorem of corresponding states}, introduced auxiliary quantities that allowed the equations for moving systems to be cast in the same form as those for systems at rest in the ether.%

In 1895, as part of a first-order treatment, Lorentz introduced the \emph{local time}, a mathematical device without physical interpretation in his theory, defined by \cite{Lorentz1895}:
\begin{equation} \label{eq14}
t' = t - \frac{x v}{c^2},
\end{equation}
where $t'$ is the local time in the moving frame, $t$ the time coordinate in the ether frame, $x$ the position coordinate, $v$ the velocity of the moving frame relative to the ether, and $c$ the speed of light. 
This local time simplified the transformation properties of the electromagnetic fields in moving bodies. It was instrumental in recovering Fresnel’s coefficient for the partial dragging of light in moving media (see the derivation in \cite{Wein1}).

First-order ether drift experiments were designed to detect small first-order effects of Earth's motion relative to the ether by measuring changes in the speed of light. James Clerk Maxwell initially suggested that measuring the speed of light in different directions could reveal Earth's motion through the ether. Still, he acknowledged that the expected effect would be too small to detect. Albert Michelson’s early experiments (1881, Potsdam) attempted to measure ether drift using interferometry \cite{Mic}. 
Michelson’s decisive innovation was not merely another attempt to detect ether drift, but the introduction of a new class of optical instrument---the interferometer---deliberately designed to register second-order effects of order $\dfrac{v^{2}}{c^{2}}$ after first-order tests had failed. 
Earlier ether-drift experiments (Arago's prism and Fizeau's water tube experiment) tested effects linear in $\dfrac{v}{c}$, and their null or Fresnel-compatible outcomes could still be reconciled with partial ether-drag hypotheses. 

Michelson saw, following Maxwell’s suggestion, that if the ether is truly stationary, then a light-speed anisotropy of order $\dfrac{v^{2}}{c^{2}}$ ought to be detectable by sending light back and forth along perpendicular arms and recombining it interferentially. He therefore devised an instrument in which the measurable quantity is a fringe displacement proportional to the difference of optical paths along two orthogonal directions, doubling the path by multiple reflections to amplify the second-order term. 

This conceptual shift---from measuring light speeds to measuring interference phase as a proxy for second-order anisotropy---provided the first experimental architecture capable, in principle, of deciding between stationary-ether models and the surviving ether-drag theories left viable by first-order nulls. 

Rather than the 1881 or 1887 runs, the instrument itself was the invention of consequence. It defined a new standard of precision physics, in which the Michelson-Morley result became not just another null result, but the first high-sensitivity second-order discrimination test of the ether concept.

To reduce vibrations, Michelson and Edward Morley refined Michelson's interferometer with a floating stone base in a mercury bath. They expected to observe an interference fringe shift due to the Earth's motion through the ether and found a change well below their expectations. They published their results in 1887, concluding that no significant ether drift was detected, contradicting the hypothesis of a stationary ether. These experiments collectively contradicted expectations from Lorentz’s ether-based electron theory \cite{MM}. 

Reflecting on the Michelson-Morley null result, Lorentz suggested a new hypothesis to support the stationary ether, namely, contraction. He wrote \cite{Lorentz1892}:

\begin{quote}
"I found only one way to reconcile the result with Fresnel’s theory. It consists of the assumption that the line joining two points of a solid body doesn’t conserve its length when it is once in motion parallel to the Earth’s motion and afterwards normal to it. \ldots\ Such a change in length \ldots\ is really not inconceivable as it seems to me."
\end{quote}

In sections \S90–\S91 of his \textit{Versuch}, Lorentz addressed the problem posed by the null result of the Michelson–Morley experiment within the framework of the stationary ether theory. He sought to remove the contradiction between Fresnel’s theory (see section \ref{FR}) and Michelson’s result. Lorentz wrote \cite{Lorentz1895}:%
\footnote{Lorentz adds footnote number 2 on p. 122 of the \emph{Versuch} \cite{Lorentz1895}:

\begin{quote}
As Mr. FitzGerald kindly informed me, he has addressed this hypothesis in his lectures for some time. In the literature, I have found it mentioned only by Mr. Lodge, in the paper "Aberration Problems" (London Philosophical Transactions, Vol. 184, A, p. 727, 1893). 
\end{quote}
immediately after introducing the contraction hypothesis in section \S 90.
He acknowledges that FitzGerald had independently proposed (in 1889) that motion through the ether could cause a physical contraction of material bodies, accounting for the Michelson–Morley null result.}

\begin{quote}
Indeed, this can be achieved through a hypothesis I expressed some time ago, and to which, as I later learned, Mr. Fitzgerald has also arrived.    
\end{quote}

To that end, Lorentz suggested the idea that the dimensions of the apparatus changed oppositely, so that when one arm of the interferometer was aligned with the ether wind, it became slightly shorter, and when perpendicular, somewhat longer. In this way, the phase differences could be mutually compensated. This reasoning led Lorentz to postulate that the \textit{body's motion through the ether physically alters its dimensions}.

In section §91, under the heading "Change of Dimensions through Translation" (\textit{Änderung der Dimensionen durch die Translation}), Lorentz presented this hypothesis explicitly. 

Lorentz assumes one arm $P$ of the interferometer lies parallel to the Earth's motion through the ether (with velocity $v$), while the other arm $Q$ is perpendicular. He denotes by $L$ the one-way length of each arm, so that the physical round-trip distance is $2L$. The predicted round-trip times in a stationary ether are as follows:

\begin{enumerate}
    \item For the arm $P$ aligned with the motion: 

\begin{equation} \label{Tpara}
T_{\parallel}
= \frac{L}{c-v} + \frac{L}{c+v}
= \frac{2Lc}{c^{2}-v^{2}}
= \frac{2L}{c}\,\frac{1}{1-\frac{v^{2}}{c^{2}}}.
\end{equation}
The first–order Taylor approximation of \eqref{Tpara} is of an inherently second–order quantity:

\begin{equation}
T_{\parallel}
=\frac{2L}{c}\,\frac{1}{1-\boxed{\frac{v^{2}}{c^{2}}}}
=\frac{2L}{c}\!\left(1+\frac{v^{2}}{c^{2}}+\cdots\right).
\end{equation}
The correction is multiples of $\frac{v^2}{c^2}$, not $\frac{v}{c}$.
Thus, for the arm aligned with the motion, one finds:

\begin{equation} \label{Tpar}
T_{\parallel}
= \frac{2L}{c}\cdot \frac{1}{1-\frac{v^{2}}{c^{2}}}
\;\approx\;
\frac{2L}{c}\!\left(1+\frac{v^{2}}{c^{2}}\right).
\end{equation}

\item For the perpendicular arm $Q$:
We do a Pythagorean construction, leading to the round-trip time:
\begin{equation}
T_\perp
= \frac{2L}{\sqrt{c^2 - v^2}}
= \frac{2L}{c}\,\frac{1}{\sqrt{1-\frac{v^{2}}{c^{2}}}}.    
\end{equation}
Again, we take the first–order Taylor approximation of this equation: 
\begin{equation}
T_{\perp}
=\frac{2L}{c}\,\frac{1}{\sqrt{1-\boxed{\frac{v^{2}}{c^{2}}}}}
=\frac{2L}{c}\!\left(1+\frac{v^{2}}{2c^{2}}+\cdots\right),
\end{equation} 

and keep only the leading correction term, and that correction is of order $\dfrac{v^2}{c^2}$:

\begin{equation} \label{Tperp}
T_{\perp} =\frac{2L}{c}\, \cdot \frac{1}{\sqrt{1-\frac{v^{2}}{c^{2}}}}
\approx \frac{2L}{c}\!\left(1+\frac{v^{2}}{2c^{2}}\right).
\end{equation}

\end{enumerate}
Therefore, the predicted second-order time difference is \eqref{Tpar} $-$ \eqref{Tperp}:
\begin{equation} \label{deltaT}
\Delta T \;=\; T_{\parallel}-T_{\perp}
\;=\; \frac{2L}{c}\left(\frac{v^{2}}{2c^{2}}\right).
\end{equation}
This is the second-order effect Michelson expected to detect.

From his calculation of the round–trip times in a stationary ether, Lorentz finds that (if dimensions remained unchanged) the arm parallel to motion would require more time than the perpendicular one by an amount proportional to a second–order time difference \eqref{deltaT} \cite{Lorentz1895}:
\begin{equation} \label{deltaT-1}
\frac{L\,v^{2}}{c^{3}}.
\end{equation}
So, if nothing else changed, Michelson should have seen a fringe shift, a second–order phase difference. However, he did not observe this.

Lorentz computed that due to motion through the ether, the two arms of the interferometer would require different round-trip times. 
Michelson's interferometer compares the round-trip travel times of light along two orthogonal
arms of equal rest length.  Lorentz now \emph{hypothesizes} that motion through the ether produces
a real physical contraction of the arm parallel to the motion, so that the two arms differ in length
by an amount $\Delta L$.  Such a physical shortening would by itself produce a round-trip delay:
\begin{equation}\label{deltag}
\Delta T_{\rm geom} = \frac{2\,\Delta L}{c},
\end{equation}
with $c$ being the speed of light in the ether frame.  Lorentz then requires that this geometrical delay
exactly compensate for the ether--drift time difference.

\noindent Equating \eqref{deltag} $\Delta T_{\rm geom}=\Delta T_{\rm ether}$ \eqref{deltaT-1} gives:
\begin{equation} \label{deltaL-2}
\frac{2\,\Delta L}{c} = \frac{L\,v^{2}}{c^{3}}
\quad\Rightarrow\quad
\Delta L = \frac{L}{2}\frac{v^{2}}{c^{2}}.
\end{equation}
Thus, the difference is of order \eqref{deltaL-2} \cite{Lorentz1895}:

\begin{equation}
\frac{L}{2}\frac{v^{2}}{c^{2}}.    
\end{equation}
Lorentz then introduces hypothetical deformations. Let the arm parallel to motion change its length by a fractional amount $\delta$ and the perpendicular arm by $\varepsilon$. Lorentz writes that the difference between the transverse deformation and the longitudinal deformation must be precisely such that it cancels the optical second-order effect. The geometric deformation must oppose, i.e., compensate for the optical effect. Thus, the longitudinal dimension contracts while the transverse does not: $\delta < \varepsilon$. 
Then, the change in arm lengths contributes an additional geometric time difference of an order:
\begin{equation} \label{deltag-2}
\Delta T_{\rm geom} = (\delta -\varepsilon) \frac{2\,L}{c},
\end{equation}
where $\Delta \equiv \delta -\varepsilon$ in \eqref{deltag}.
This is a geometric light-path delay produced by arm deformation, inserted to compensate for Michelson’s predicted drift signal.

\noindent Now, Lorentz imposes the compensation condition:
\begin{equation}
\underbrace{\Delta T_{\text{ether}}}_{\text{predicted from ether drift \eqref{deltaT-1}}}
\;+\;
\underbrace{\Delta T_{\text{geom}}}_{\text{due to changed arm lengths \eqref{deltag-2}}}
=0.
\end{equation}
Setting the sum to zero gives:
\begin{equation}
(\delta -\varepsilon)\,\frac{2L}{c} + \frac{L v^{2}}{c^{3}} = 0.
\end{equation}
Dividing both sides by $\dfrac{L}{c}$ yields:
\begin{equation}
2(\delta-\varepsilon) + \frac{v^{2}}{c^{2}} = 0,
\end{equation}
the amount required to nullify the expected second-order anisotropy \cite{Lorentz1895}:
\begin{equation}\label{eq:119}
\delta-\varepsilon \;=\; -\frac{v^{2}}{2c^{2}}, \qquad \text{The compensation condition.}
\end{equation}
This is the compensating requirement.

In section \S 92, Lorentz adds a physical plausibility argument. In an equilibrium configuration under translation, only the longitudinal dimension should change $(\varepsilon=0)$, while the transverse dimension remains unaltered. Substituting the physical hypothesis $\varepsilon=0$ into \eqref{eq:119} yields:
\begin{equation} \label{deltav}
\delta \;=\; -\frac{v^{2}}{2c^{2}}.
\end{equation}
Thus, Lorentz concludes that the contraction would produce a shortening in the direction of motion in the ratio of \cite{Lorentz1895}:%
\footnote{The following is a modern reconstruction of Lorentz’s reasoning, not a verbatim transcription of his text. We introduce the fractional deformations:
\begin{equation} \label{parallel}
L_\parallel = L(1+\delta), \qquad L_\perp = L(1+\varepsilon).
\end{equation}
Imposing Lorentz’s hypothesis $\varepsilon=0$ and substituting $\delta$ from \eqref{deltav} gives:
\begin{equation}
L_\parallel = L\!\left(1-\frac{v^{2}}{2c^{2}}\right).
\end{equation}
But this is precisely the second-order Taylor expansion of the exact square-root factor:
\begin{equation}
L_\parallel = L\sqrt{1-\frac{v^{2}}{c^{2}}}
= L\left(1-\frac{v^{2}}{2c^{2}} + O\!\left(\frac{v^{4}}{c^{4}}\right)\right).
\end{equation}}

\begin{equation} \label{factor}
1\, : \, \sqrt{1-\frac{v^{2}}{c^{2}}}, 
\end{equation}
which is precisely the shortening required to suppress the Michelson effect. 

Importantly, this exact square-root form \eqref{factor} appears only after Lorentz introduces in \S92 the \emph{equilibrium-of-molecular-forces} argument, and after the two conditions $\varepsilon=0$ and \eqref{deltav} are combined. In sections \S90--91, one obtains merely the \emph{second-order} compensation condition; the \emph{full exact factor} $\sqrt{1-\frac{v^{2}}{c^{2}}}$ first enters through the physical assumption of molecular equilibrium under translation.

Lorentz explicitly grounds this step in a physical rationale. However paradoxical the hypothesis may initially seem, it is, he argues, not implausible if one assumes that molecular forces are transmitted through the ether in the same manner as electric and magnetic interactions. If this is the case, then uniform translation through the ether will necessarily modify the interaction between molecules or atoms, just as it modifies the forces between charged bodies in motion. Since the equilibrium shape and dimensions of a solid body depend on the intensity of these molecular forces, any such modification must, in turn, produce a corresponding change in the body’s dimensions.

Theoretically, Lorentz argued that the hypothesis could not be rejected. If molecular forces are transmitted through the ether, translation through the ether should modify those forces and alter the body dimensions. For the Earth’s orbital velocity, the resulting deformation is of second order \eqref{108}. One diameter of the Earth would shorten by roughly $6.5\,\mathrm{cm}$, and a one-meter rod would change by about $10^{-4}\,\mu\mathrm{m}$ when rotated between principal orientations. Only an interference method could ever hope to detect effects of this magnitude. Under such conditions, repeating the Michelson experiment would again reveal no fringe displacement. The \emph{null result} is explained because the contraction of the arms compensates exactly the optical path difference predicted by Maxwell’s theory \cite{Lorentz1895}. 

This was the first coherent formulation of what later became known as the Lorentz–FitzGerald contraction—an effect that Einstein would not treat dynamically but re-interpret kinematically under the relativity principle.
Lorentz did not present the contraction as something to be detected directly. To look for the contraction itself would undermine the very role for which it was introduced. It was posited to eliminate the ether-drift signature that interferometry was expected to reveal. The contraction was framed as a compensating dynamical effect of ether-mediated molecular forces, tuned to cancel the anticipated second-order optical anisotropy. It is not an independent observable, but rather a mechanism that ensures a null result in Michelson-type experiments. This is a teleological non-observability: The contraction's function is to hide the effect that would otherwise be detectable.

\subsection{The Empire of Contradictions}

The tangle of experiments suggested that either:  

1. \emph{The Earth was moving through a stationary ether} (the Lorentz-Fresnel hypothesis): This was unlikely. The Michelson–Morley experiment suggests Earth carries ether with it; otherwise, an ether wind should be detectable. 

2. \emph{The ether was somehow dragged along with Earth} (Stokes' hypothesis): The ether is carried along completely by the Earth, so there is no relative motion between Earth and ether. 

However, full ether drag was also unlikely as it conflicted with Fresnel’s formula \eqref{eq21}, which describes light in a transparent medium. It explained Fizeau’s experiment with moving water. 

Full ether drag also conflicted with stellar aberration observations (see section \ref{BA}). If Earth were to drag the ether fully, then near Earth, there would be no relative motion between Earth and the ether. In that case, starlight would propagate straight down the telescope, independent of Earth’s orbital motion.
However, in reality, we observe aberration; the telescope must be tilted slightly to catch the light, as if the Earth were moving through a stationary ether.
Therefore, full ether drag is incompatible with the principle of stellar aberration.

3. \emph{There was no ether at all}: This implied that the speed of light was constant in all directions. This result was consistent with the Michelson-Morley experiment, Fresnel's formula, and stellar aberration, but was inconsistent with the Newtonian law of addition of velocities. Still, it was in line with Maxwell’s prediction that light should always travel at the speed of light $c$ in a vacuum, independent of the source’s motion.

If we imagine full ether drag, the Michelson–Morley experiment is trivially explained; however, aberration disappears.
If we retain Fresnel’s partial drag, we avoid aberration and explain Fizeau's water tube experiment; however, the result of the Michelson–Morley experiment then requires contraction. To Lorentz, that patch seemed satisfactory. The ether remained stationary, the Fresnel partial drag explained the Fizeau water tube experiment, and contraction explained the Michelson–Morley experiment. The contraction was a physical deformation of bodies, caused by the interaction of their intermolecular forces with the ether. Thus, Newtonian kinematics was left untouched, and the strangeness was shifted into the physics of molecular forces. By treating contraction as a property of matter, Lorentz avoided confronting the fact that it implied a more profound inconsistency between Galilean kinematics and Maxwellian electrodynamics. But in hindsight, this was a fragile balancing act because each phenomenon had to be explained by a special \emph{ad hoc} hypothesis. Lorentz and others did not recognize that the whole pattern of contradictions pointed to the failure of Newtonian kinematics itself.

\section{Einstein's Path to Special Relativity}

\subsection{A Frozen Light} \label{C}

Einstein undoubtedly covered and discarded countless scraps of paper, including patent drafts, while working through problems in electrodynamics. Those ideas also accompanied him on long walks home with Michele Besso. Yet, despite years of preoccupation with the subject, Einstein published nothing on optics or electrodynamics before 1905. Until otherwise discovered, no notebooks or drafts from that formative period have been found to survive. Historians, therefore, face the task of reconstructing the route to special relativity from scattered and fragmentary traces. Contradictions compound the difficulty. Later, Einstein occasionally answered questions about his development, but offered divergent accounts at different times — not necessarily due to inconsistencies in his thought, but rather from shifting contexts and the inevitable selectivity of recollection. Thus, we cannot assemble a coherent picture of his pathway to special relativity. The following reconstruction is based on available sources, including surviving publications, correspondence, retrospective testimonies, and the broader intellectual milieu in which Einstein worked.

\medskip

In Newtonian (Galilean) mechanics, velocities transform by simple addition or subtraction. If an object moves at velocity $v$ relative to some rest frame, and an observer moves at velocity $u$ in the same direction, then 
the observer measures the object's speed as $v-u$. For everyday objects, this rule works well. For instance, if a train travels at $150\,\mathrm{km/h}$ and we run alongside it at $10\,\mathrm{km/h}$, we perceive the train to move at $140\,\mathrm{km/h}$ relative to us. Applied naively to light, this would suggest that an observer moving at $c$ in the same direction as a light beam (also traveling at $c$) should measure its relative speed as $c-c = 0$. In other words, the beam should appear stationary, a "frozen" oscillation.  

Maxwell’s theory requires light to consist of time-varying electric and magnetic fields that perpetually regenerate one another. A wave "at rest" would collapse; it could not exist. Moreover, Maxwell’s equations predict that the speed of light is fixed at c, independent of the source’s motion. A stationary, "frozen" light wave is impossible in Maxwellian electrodynamics for these two reasons:

\begin{enumerate}
    \item According to Maxwell’s equations, electromagnetic waves are governed by:
\begin{equation} \label{waves}
\nabla^2 \mathbf{E} - \boxed{\frac{1}{c^2}}\frac{\partial^2 \mathbf{E}}{\partial t^2} = 0, 
\qquad 
\nabla^2 \mathbf{B} - \boxed{\frac{1}{c^2}}\frac{\partial^2 \mathbf{B}}{\partial t^2} = 0 .
\end{equation}
These wave equations have a fixed propagation speed:
\begin{equation} 
c = \frac{1}{\sqrt{\mu_0 \epsilon_0}},
\end{equation}
independent of the motion of the source. Thus, unlike in Galilean kinematics, one cannot subtract the observer’s velocity from $c$ to make the wave stand still. 

\item Moreover, the very structure of Maxwell’s equations 
requires the electric and magnetic fields to oscillate in time: 
\begin{equation} \label{Faraday}
\nabla \times \mathbf{E} = -\frac{\partial \mathbf{B}}{\partial t},  \qquad \qquad \text{(Faraday's law)} \quad \text{and:}
\end{equation}
\begin{equation} \label{Ampere}
\nabla \times \mathbf{B} = \mu_0\epsilon_0 \frac{\partial \mathbf{E}}{\partial t} \qquad \text{(Ampére--Maxwell law)}.     
\end{equation}
 
\noindent Faraday’s law says that a magnetic field that changes in time induces a curl in $\mathbf{E}$, and the Ampère–Maxwell law says that an electric field that changes in time induces a curl in $\mathbf{B}$.
Maxwell’s curl equations couple the electric $\mathbf{E}$ and magnetic $\mathbf{B}$ fields.

Suppose, for a moment, that both fields are static:  

\begin{equation}
\frac{\partial \mathbf{B}}{\partial t} = 0, \qquad \frac{\partial \mathbf{E}}{\partial t} = 0.    
\end{equation}
Then the curl equations reduce to:

\begin{equation} 
\nabla \times \mathbf{E} = 0 
\quad \text{(Faraday's law)}, \quad \nabla \times \mathbf{B} = 0 
\quad \text{(Ampére--Maxwell law)}.
\end{equation}
Both fields are irrotational in a vacuum (no charges, no currents).
We combine that with the other vacuum Maxwell equations:
\begin{equation} \label{Gauss}
\nabla\cdot\mathbf{E}=0, \qquad \nabla\cdot\mathbf{B}=0 \qquad \text{(Gauss laws)},
\end{equation}
and the only solution consistent with all four is:  

\begin{equation}
\mathbf{E} = \mathbf{B} = 0.
\end{equation}
That is, static, source-free solutions must be trivial.
Hence, any nonzero solution requires time-varying fields. 
If the time-dependence vanished, the fields would disappear altogether. Thus, this argument is a no-go theorem for frozen light in a vacuum. If the fields stop evolving, Maxwell’s equations say the fields must collapse to zero. There is no static, non-trivial configuration in a vacuum that carries energy and momentum without oscillating.

\end{enumerate}

\subsection{The Chasing-a-Light-Beam Episode as Retrospective Narrative}

Einstein would later describe the above conflict between mechanics and electrodynamics as his vision of \emph{chasing a beam of light}  \cite{Einstein49, Seelig1956, Wertheimer1916}.

John Norton argues that the familiar scene of chasing a light beam was not a formative step in Einstein’s path to special relativity, but rather a later, retrospective origin story. In his reading, the anecdote crystallized only when interviewers and biographers pressed Einstein for a simple point of departure, a memorable image of a boy running after a light ray. The tale satisfied a public appetite for poetic beginnings, and Einstein eventually reproduced it himself \cite{Nor1}.

Norton juxtaposes two sources: Einstein’s 1946 \emph{Autobiographical Notes} \cite{Einstein49} and an earlier recollection from his student days in Aarau \cite{Seelig1956}. In the 1946 text, Einstein reports that if one pursued a light beam at the speed of light $c$, one would see a frozen electromagnetic wave — forbidden by Maxwell’s equations — and calls this the germ of the special theory of relativity, framing it as the starting point of his lifelong search for a universal principle analogous to thermodynamics. Norton treats this as a rational reconstruction, shaped by hindsight. By contrast, in the earlier Aarau reminiscence, Einstein refers to it as a childlike thought experiment that concludes only in puzzlement over a time-independent wave field, with no connection to relativity, invariance, or any other principles. 

Norton further proposes that the 1946 version of the thought experiment was influenced by Max Wertheimer, the Gestalt psychologist who interviewed Einstein in 1916 and, in 1945, published his account in \emph{Productive Thinking} \cite{Wertheimer1916}. Wertheimer sent Einstein drafts shortly before his death, and Norton suggests that seeing his own narrative in Wertheimer’s framing may have prompted Einstein to fold the light-beam anecdote into the \emph{Autobiographical Notes}. On this reading, Einstein may have recovered or reconstructed the memory, blending youthful impressions with later insights into electrodynamics. When he invoked Maxwell’s equations, Norton suggests, he was projecting mature knowledge backward, remembering the adolescent puzzle through the lens of the seasoned scientist \cite{Nor1}.

Norton offers a more coherent reinterpretation of the puzzle itself. The notion of frozen light becomes intelligible when connected to Einstein’s later work on emission theories of light. Recall that if light were emitted by moving sources, and if the classical Galilean additional law of velocities held, then light from a source moving at $c$ relative to the observer should also move at $c-c = 0$, forming a frozen wave — something never observed. This reasoning, Norton notes, aligns with Einstein’s later critique of emission theories. He therefore speculates that Einstein, decades later, unconsciously merged these mature reflections with his vague adolescent image, creating a hybrid, retrospective narrative \cite{Nor1}.

Yet Norton’s reconstruction, as compelling as it sounds, is highly speculative. No archival evidence supports it. Indeed, no letters, drafts, or notebooks show Einstein inventing or adapting the story of chasing a light beam. Norton’s main argument for a late construction is the absence of early mention, but silence in the record does not prove nonexistence. Einstein often omitted key conceptual steps from his publications and simplified narratives for different audiences. Thus, turning absence into proof of invention is methodologically precarious. 

Norton’s broader project, however, is less biographical than methodological. He reacts against the hagiographic tendency to treat Einstein’s recollections as transparent truth. His goal is to de-mythologize, i.e., to remind historians that autobiographical memory, especially after fifty years, is selective and shaped by circumstance. In that sense, his skepticism is salutary. 

Methodologically, Norton follows the interpretive-reconstructive approach, seeking psychologically plausible narratives to bridge gaps in the record. This contrasts with the documentarian tradition of scholars like John Stachel and Jürgen Renn, who ground their analyses in written evidence such as correspondence and Einstein's later manuscripts and publications. For example, Stachel’s analysis of Einstein’s route to special relativity rests on the unpublished 1912 manuscript, where Einstein discusses emission theories of light \cite{Einstein12}. 

Each method has strengths and weaknesses. The documentarian approach cannot access unrecorded thought, while the reconstructive approach risks conjecture and narrative inflation. Norton’s choice of the latter inevitably introduces a new myth, that of the rational historian who sees through Einstein’s self-presentation.

Ultimately, Norton’s scenario remains unverified. His psychological inferences—whether Einstein embellished the light-beam story or recalled it under suggestion—are untestable. Another historian could build an equally coherent but contrary interpretation, that Einstein’s recollection was genuine though simplified, or that the youthful idea was embryonic but real. Persuasion here depends not on proof but on narrative coherence.

Even so, Norton’s skepticism performs a valuable corrective function. It reminds readers that Einstein’s public persona was shaped gradually, that his recollections were mediated by decades of retelling, and that his actual intellectual labor lay in the sustained theoretical work on electrodynamics, not in single flashes of insight. 
Thus, while Norton cannot demonstrate that Einstein embellished the story of chasing a light beam, he successfully shifts attention from mythic moments to the protracted process of reasoning and reconstruction. In that modest sense, his critique stands. The episode should not be taken literally as the genesis of relativity but rather as a symbolic condensation of a long, intricate intellectual journey.

\subsection{The Light Beam Paradox}

This was the paradox Einstein faced. Maxwell’s equations were not Galilean-invariant. In Galilean mechanics, we could write a transformed velocity:
$v' = v - u$, and so if $v=c$ and $u=c$, we would get $v'=0$. But according to Maxwell’s equations, the wave equation fixes the form of light propagation to always be at $c$, independent of the motion of the source. In other words, the electromagnetic field equations are not invariant under Galilean transformations.

Consider a Galilean boost with constant velocity $\mathbf{u}$:
\begin{equation} \label{Gal}
t' = t, \qquad \mathbf{r}' = \mathbf{r} - \mathbf{u}\,t \qquad \text{or} \qquad \mathbf{r} = \mathbf{r'} + \mathbf{u}\,t'. 
\end{equation}
For any field $F$, $F'(\mathbf{r}',t')\equiv F(\mathbf{r},t)$, evaluated at the same event, the spatial derivatives are unchanged:%
\footnote{For any field $F$, we write: $F'(\mathbf{r}',t')\equiv F(\mathbf{r},t)=F(\mathbf{r'} + \mathbf{u}t',t')$ evaluated at the same event. If we hold $t$ fixed, then $d \mathbf{r} = d \mathbf{r}'$, since $\mathbf{r} = \mathbf{r'} + \mathbf{u}\,t'$, and $d t' = 0$. Thus, the spatial derivatives are unchanged: $\nabla' F'(\mathbf{r}',t')  = \nabla F(\mathbf{r},t)$, where:

\begin{equation} \label{nab}
\nabla F = \left( \frac{\partial F}{\partial r_1}, \frac{\partial F}{\partial r_2}, \frac{\partial F}{\partial r_3} \right).    
\end{equation}
Because this holds for any $F$, it is the operator identity \eqref{nabla}.}

\begin{equation} \label{nabla}
\nabla' = \nabla.
\end{equation}
However, due to the chain rule, the time derivative picks up an extra convective term:%
\footnote{Now we hold $\mathbf{r}'$ fixed and differentiate $F'(\mathbf{r}',t')$ with respect to $t'$, using the chain rule: 

\begin{equation}
\frac{\partial F'}{\partial t'}
=\frac{\partial F}{\partial t}\frac{\partial t}{\partial t'}
+\sum_{_i}^3 \frac{\partial F}{\partial r_i}\frac{\partial r_i}{\partial t'}= \frac{\partial F}{\partial t}\frac{\partial t}{\partial t'} +\frac{\partial \mathbf{r}}{\partial t'} \cdot\nabla F.
\end{equation}
Recognizing that the sum becomes a dot product \eqref{nab} and 
$\dfrac{\partial t}{\partial t'} =1, \, \dfrac{\partial \mathbf{r}}{\partial t'}= \mathbf{u}$, we get:

\begin{equation}
\frac{\partial F'}{\partial t'}
=\frac{\partial F}{\partial t}
+\left(\frac{\partial \mathbf{r}}{\partial t'}\cdot\nabla F\right)
=\frac{\partial F}{\partial t}+\mathbf{u}\cdot\nabla F.
\end{equation}
Since this holds for any $F$, as operators, we see that the time derivative picks up an extra convective term \eqref{chain}.}

\begin{equation} \label{chain}
\frac{\partial}{\partial t} = \frac{\partial}{\partial t'} - \mathbf{u}\cdot\nabla' .
\end{equation}
We take Maxwell's equations \eqref{Faraday} and \eqref{Ampere} in vacuum and \eqref{Gauss}.

\noindent We rewrite them in primed variables. Recall equation \eqref{nabla}.

\medskip

\textit{1. Faraday's law.}

\medskip

Substituting the time derivative \eqref{chain} into Faraday's law \eqref{Faraday}, we get:

\begin{equation} \label{Farchain}
\nabla'\times\mathbf{E}' \;=\; -\Big(\frac{\partial}{\partial t'} - \mathbf{u}\cdot\nabla'\Big)\mathbf{B}'
\;=\; -\,\frac{\partial\mathbf{B}'}{\partial t'} + \big(\mathbf{u}\cdot\nabla'\big)\mathbf{B}' .
\end{equation}
However, the desired primed form is: 
\begin{equation} \label{primed}
\nabla'\times\mathbf{E}'=-\,\partial\mathbf{B}'/\partial t'.    
\end{equation}
Comparing \eqref{Farchain} with the desired primed form \eqref{primed}, we find there is an extra term
$+\,(\mathbf{u}\cdot\nabla')\mathbf{B}'$ 
compared with the desired primed form \eqref{primed}.

\medskip

\textit{2. Ampére--Maxwell's law}.

\medskip

Substituting the time derivative \eqref{chain} into Ampére's law \eqref{Ampere}, we get:
\begin{equation} \label{Amptra}
\nabla'\times\mathbf{B}' \;=\; \mu_0\varepsilon_0\Big(\frac{\partial}{\partial t'} - \mathbf{u}\cdot\nabla'\Big)\mathbf{E}'
\;=\; \mu_0\varepsilon_0\,\frac{\partial\mathbf{E}'}{\partial t'} - \mu_0\varepsilon_0\big(\mathbf{u}\cdot\nabla'\big)\mathbf{E}' .
\end{equation}
Again, relative to the desired primed form:
\begin{equation} \label{primed2}
\nabla'\times\mathbf{B}'=\mu_0\varepsilon_0\,\partial\mathbf{E}'/\partial t',    
\end{equation}
in equation \eqref{Amptra}, there is an extra term 
$$-\mu_0\varepsilon_0(\mathbf{u}\cdot\nabla')\mathbf{E}'.$$

\textit{3. Gauss laws}. The Gauss laws \eqref{Gauss} keep their form under the Galilean transformations:
\begin{equation}
\nabla'\cdot\mathbf{E}'=0, \qquad \nabla'\cdot\mathbf{B}'=0,
\end{equation}
but the curl equations do not. Hence, Maxwell's equations are not invariant under Galilean transformations.

\medskip

We can also see it at the wave-equation level. Consider the unprimed wave equations \eqref{waves}.
Substituting the time derivative \eqref{chain} into equations \eqref{waves} we get equation \eqref{waves} in primed variables:%
\footnote{Starting from the vacuum wave equations \eqref{waves}, a Galilean boost with \eqref{Gal} implies (via the chain rule) \eqref{nabla} and \eqref{chain}. Therefore, each wave equation separately becomes, in primed variables:
\begin{equation} \label{E'}
\Big[\nabla'^2-\frac{1}{c^2}\Big(\frac{\partial}{\partial t'}-\mathbf{u}\cdot\nabla'\Big)^2\Big]\mathbf{E}'=\mathbf{0},
\end{equation}

\begin{equation} \label{B'}
\Big[\nabla'^2-\frac{1}{c^2}\Big(\frac{\partial}{\partial t'}-\mathbf{u}\cdot\nabla'\Big)^2\Big]\mathbf{B}'=\mathbf{0}.
\end{equation}
The compact operator \eqref{compact} covers both original wave equations \eqref{E'} and \eqref{B'}. We apply it to $E'$ and $B'$ (component by component). Expanding the square makes the non-invariance explicit,
where the mixed term and the $(\mathbf{u}\cdot\nabla')^2$ piece are exactly what spoil Galilean invariance.}

\begin{equation} \label{compact}
\Big[\nabla'^2 - \frac{1}{c^2}\Big(\frac{\partial}{\partial t'} - \mathbf{u}\cdot\nabla'\Big)^2\Big]\mathbf{E}', \mathbf{B}' = \mathbf{0},
\end{equation}
which expands to
\begin{equation}
\Big(\nabla'^2-\frac{1}{c^2}\frac{\partial^2}{\partial t'^2}
+\frac{2}{c^2}(\mathbf{u}\cdot\nabla')\frac{\partial}{\partial t'}
-\frac{1}{c^2}(\mathbf{u}\cdot\nabla')^2\Big){\mathbf{E}',\mathbf{B}'}=\mathbf{0}.
\end{equation}
The extra mixed term in this equation and the $$(\mathbf{u}\cdot\nabla')^2,$$ term mean the equation’s form is not preserved under Galilean transformations \eqref{Gal}.

\subsection{Faraday’s Induction} \label{Fa}

Einstein’s 1905 paper on special relativity \cite{Einstein05} opens with the observation that classical electrodynamics asymmetrically treats Faraday’s induction experiment. Faraday’s law states that a changing magnetic field induces an electric current in a conductor. Thus, when a magnet and a conductor undergo relative motion, a current is induced in the conductor. In the ether theory, however, the explanation depends on which body is assumed to move through the ether. 

If the magnet moves while the conductor is at rest in the ether, Faraday’s law \eqref{Faraday} implies that the time-varying field induces an electric field in the ether, which in turn drives the current in the conductor. But if the conductor moves and the magnet is stationary in the ether, no such induced electric field exists; instead, the current is explained by the Lorentz force acting on the moving charges:

\begin{equation} \label{eq22}
\mathbf{F} = q (\mathbf{v} \times \mathbf{B}).
\end{equation}

Einstein noted that this explanatory asymmetry is purely artificial. The observable outcome, an induced current, depends solely on the relative motion of magnet and conductor, not on which one is taken to move relative to an ether. Therefore, he argued, the electrodynamics of moving bodies should not depend on an absolute ether frame, but must instead be formulated on a principle of relativity.

In 1952, Einstein reflected on his early thoughts that led to the development of special relativity. He stated that his motivation was primarily the realization that the electromotive force induced in a moving conductor is nothing but an electric field. This reinforced his view that Maxwell’s equations should hold in all inertial frames, leading him to the principle of relativity \cite{Nor}. Max Born, who frequently interacted with Einstein in 1915, recalled that Faraday’s induction law had a greater impact on Einstein’s thinking than the Michelson-Morley experiment. 
This law had been well known for decades and was understood to depend only on relative motion; yet, its explanation was inconsistent in classical electrodynamics \cite{Born1}, \cite{Born2}. 
In his 1920 manuscript, Einstein explicitly stated that the magnet and conductor experiment played a crucial role in the development of special relativity. The theoretical inconsistency between Faraday’s induction and Maxwell-Lorentz theory forced him to reconsider the foundations of physics \cite{Einstein20}. 

Trying to transform the electromagnetic equations using the Galilean transformation \eqref{Gal}, the equations change their form and no longer describe the same physics.
One might try to rescue invariance by allowing $\mathbf{E}$ and $\mathbf{B}$ to mix under the Galilean transformation. In the Galilean physics, the electric field mixes with the magnetic field in the following way:

\begin{equation} \label{GT}
\mathbf{E}'=\mathbf{E}+\mathbf{u}\times\mathbf{B},\qquad
\mathbf{B}'=\mathbf{B}-\frac{1}{c^2}\,\mathbf{u}\times\mathbf{E},
\end{equation}
These are the so-called “Galilean” limit used in pre-relativistic electrodynamics, i.e., the $\dfrac{u}{c} \ll1$ limit of the Lorentz transformations.

Let us test Faraday's law in a vacuum.
We compute the left-hand side using the so-called "Galilean" transformations for the electric and magnetic fields \eqref{GT}, the identity \eqref{nabla}, and the constant $\mathbf{u}$:%
\footnote{We use the field equation $\mathbf{E}'$ \eqref{GT} and \eqref{nabla}:

\begin{equation} \label{nabE'}
\nabla'\times\mathbf{E}'
= \nabla\times\big(\mathbf{E}+\mathbf{u}\times\mathbf{B}\big)
= \underbrace{\nabla\times\mathbf{E}}_{=-\, \dfrac{\partial\mathbf{B}}{\partial t}}
\;+\; \nabla\times(\mathbf{u}\times\mathbf{B}).
\end{equation}
Using the vector identity:
\begin{equation}
\nabla\times(\mathbf{u}\times\mathbf{B})
= \mathbf{u}(\nabla\cdot\mathbf{B}) - (\mathbf{u}\cdot\nabla)\mathbf{B},
\end{equation}
and the Gauss law \eqref{Gauss} $\nabla\cdot\mathbf{B}=0$, equation \eqref{nabE'} becomes the transformed Faraday equations \eqref{FarGT}.}

\begin{equation} \label{FarGT}
\nabla'\times\mathbf{E}'
= -\,\frac{\partial\mathbf{B}}{\partial t}
+ \mathbf{u}\,(\nabla\cdot\mathbf{B}) - (\mathbf{u}\cdot\nabla)\mathbf{B} 
= \boxed{-\,\frac{\partial\mathbf{B}}{\partial t} - (\mathbf{u}\cdot\nabla)\mathbf{B}.}
\end{equation}

\noindent Regarding the right-hand side, using the transformation for the electric and magnetic field \eqref{GT}, we get:%
\footnote{The desired result is $$-\frac{\partial\mathbf{B}'}{\partial t'}=-\frac{\partial\mathbf{B}'}{\partial t}.$$.} 
\begin{equation} \label{LHS}
-\frac{\partial\mathbf{B}'}{\partial t}
= -\frac{\partial}{\partial t}\!\left(\mathbf{B}-\frac{1}{c^2}\,\mathbf{u}\times\mathbf{E}\right)
= \boxed{-\,\frac{\partial\mathbf{B}}{\partial t} + \frac{1}{c^2}\,\mathbf{u}\times\frac{\partial\mathbf{E}}{\partial t}.}
\end{equation}
We equate the right-hand side of equation \eqref{LHS} and the right-hand side of \eqref{FarGT} and cancel $-\,\partial\mathbf{B}/\partial t$, and get:
\begin{equation} \label{Bc2}
-(\mathbf{u}\cdot\nabla)\mathbf{B}
= \frac{1}{c^2}\,\mathbf{u}\times\frac{\partial\mathbf{E}}{\partial t}.
\end{equation}
Using Ampére--Maxwell law \eqref{Ampere} in vacuum: 
\begin{equation}
\frac{1}{c^2}\frac{\partial \mathbf{E}}{\partial t}=\nabla\times\mathbf{B},    
\end{equation}
the right-hand side of equation \eqref{Bc2} becomes:
\begin{equation}
\frac{1}{c^2}\,\mathbf{u}\times\frac{\partial\mathbf{E}}{\partial t}
= \mathbf{u}\times(\nabla\times\mathbf{B}).
\end{equation}
Thus, the condition \eqref{Bc2} reads:
\begin{equation} \label{ind}
\boxed{\;-(\mathbf{u}\cdot\nabla)\mathbf{B} = \mathbf{u}\times(\nabla\times\mathbf{B})\; ,}
\end{equation}
which is in general false because these are distinct differential operators. Therefore, even with this best available so-called "Galilean" field transformation \eqref{GT}, Faraday’s law does not keep its form. A parallel failure occurs for the Ampére--Maxwell law \eqref{Ampere}.

Einstein faced a paradox because the Galilean boost and linear $\mathbf{E},\mathbf{B}$ transformation (with absolute time) did not preserve the complete Maxwell system of equations with finite $c$. Analyzing Faraday's law \eqref{Faraday} within the framework of Maxwell’s theory, as then interpreted, seemed to preclude that the mechanical (physical) principle of relativity [Galilean transformation \eqref{Gal}] and the field transformation \eqref{FarGT} should hold for electromagnetic phenomena. According to this principle, the laws of mechanics take the same form in all inertial reference frames. In what way, then, does Maxwell’s theory differ from the Galilei-invariant theory governed by equations \eqref{Gal} and \eqref{FarGT} (a theory which is in accord with the mechanical principle of relativity)? John Stachel and Max Jammer argue that the answer lies in the presence of the Faraday induction term \eqref{ind} (the boxed equations). According to Stachel and Jammer, this term destroys the exact Galilean invariance of Maxwell's equations. Hence, the law of induction – or the magnet and conductor experiment – sets up the conflict between electrodynamics and the Galilean principle of relativity \cite{JmS}. 

\subsection{First-Order Ether Drift Experiments} \label{Fi}

As a student at the Zurich Polytechnic, Einstein was interested in ether drift experiments and appears to have designed one around 1899. According to notes taken by Jun Ishiwara, in his 1922 lecture in Kyoto, Einstein recalled wanting to verify the motion of the ether relative to Earth using an experiment involving thermocouples. He initially did not doubt the existence of the ether and aimed to measure differences in heat generated by two light beams traveling in opposite directions relative to Earth’s motion \cite{Ish}. 

In September 1899, during a visit to Aarau, Einstein conceived the idea of investigating how a body’s motion relative to the ether affects the speed of light in transparent materials. He wrote to physicist Wilhelm Wien about this, but felt his ideas were dismissed. This suggests he was considering a variation of Fizeau’s 1851 experiment with moving water \cite{Stachel2002}. Einstein's interest in detecting Earth's motion in the ether through light experiments continued, but he was unaware of prior works by Lorentz and Michelson. He aimed to solve the problem empirically but could not build his apparatus due to skepticism from his professors, notably Heinrich Weber \cite{Reiser1930}. Max Wertheimer later described how Einstein was puzzled by the idea that light might travel at different speeds in different directions and sought experimental ways to test this \cite{Wertheimer1916}. 

By late 1899, Einstein read Wien’s 1898 paper on the motion of the ether, which referenced multiple ether drift experiments, including Michelson-Morley’s. While Einstein did not mention this experiment in his early letters, he likely became aware of it by then. 

By 1902, he had studied Lorentz’s 1895 \emph{Versuch} book \cite{Lorentz1895}, which detailed the Michelson-Morley experiment \cite{CPAE1}, Doc. 130. After graduating in 1900, Einstein designed another ether drift experiment. In December 1901, he discussed his ideas on electrodynamics with his mentor Alfred Kleiner in Zurich, who encouraged him to publish them. Einstein considered writing a paper but ultimately decided not to. In a 1901 letter to Marcel Grossmann, he described a more straightforward experimental method, based on interference effects, but never followed through with publication \cite{CPAE1}, Doc. 122.

\subsection{The Michelson-Morley Experiment} 

The Michelson-Morley experiment is often portrayed in textbooks as a crucial precursor to Einstein’s special relativity, suggesting that Einstein was either directly or indirectly influenced by it. However, Einstein gave varying accounts of its influence on his thinking, sometimes acknowledging it as significant and other times dismissing its role in his development of relativity.

Gerald Holton differentiates between Einstein’s responses when asked about Michelson’s influence and his spontaneous comments about the genesis of special relativity. When volunteering explanations, Einstein consistently emphasized the role of the Fizeau water tube experiment and stellar aberration rather than Michelson’s work. Robert Millikan, a long-time associate of Michelson, promoted the narrative that Michelson laid the experimental foundations for relativity \cite{Holton1969}. 

In a 1931 event at Caltech, Millikan introduced Michelson as the experimental pioneer of relativity, and Einstein, without challenging this view, paid tribute to Michelson’s contributions. However, Michelson was skeptical of relativity and reportedly regretted that his experiments had contributed to its development. Einstein later clarified in letters that he had taken Michelson’s results for granted and that they did not directly influence his pathway to relativity. He stated that he only learned of Michelson’s work through Lorentz’s research. 

In a 1954 letter to historian Francis Garvin Davenport, Einstein reiterated that Michelson’s experiment had little influence on his thinking, as he was already convinced of the non-existence of absolute motion. Holton argues that the association of Michelson’s experiment with relativity grew over time, particularly because there were few experimental confirmations of special relativity in the years following its publication \cite{Holton1969}. 

Physicists, such as Max Planck, initially viewed Michelson’s result as the only supporting experimental evidence. Over time, a "symbiotic" connection formed between Michelson’s experiment and relativity in the didactic literature, reinforcing the myth that the former played a significant role in the latter’s discovery. 

Between 1950 and 1954, Einstein shared his views on the Michelson-Morley experiment with Robert Shankland \cite{Shan1}, \cite{Shan2}, \cite{Shan3}. In a 1950 conversation, Einstein stated that he first learned of the experiment through Lorentz's writings, but only after 1905. He emphasized that the experimental results that most influenced him were stellar aberration and Fizeau’s measurements of light speed in moving water. 

By 1952, Einstein was unsure when he had first heard of Michelson’s experiment, stating that he had assumed its results to be true but had not consciously considered them in developing special relativity \cite{Holton1988}.

\subsection{Detours: Emission Theory and Ehrenfest’s Triad} \label{D}

Einstein's later recollections suggest that before 1905, he seriously entertained an emission theory of light, where the speed of light would depend on the velocity of the light source \cite{CPAE5}, Doc. 409. However, he eventually abandoned emission theories due to mounting theoretical difficulties.
Paul Ehrenfest described three principles relevant to theories of light propagation \cite{Ehr}:

(A) The principle of the constancy of the speed of light (as in Lorentz’s electron theory).

(B) The Newtonian addition law of velocities (Galilean transformation).

(C) The relativity principle states that physical laws are the same in all inertial frames.

An emission theory would adopt (B) and (C) while rejecting (A). This meant that light’s velocity would vary depending on the motion of its source, which is in line with Newtonian mechanics. 

Before 1905, Einstein leaned toward emission theory because he assumed the Newtonian addition law of velocities was fundamental. 
Einstein was weighing two different frameworks:
1. Lorentz’s ether-based electron theory. 2. Newtonian-style emission theories. 

If Einstein had started with Newton’s kinematics (B) + the relativity principle alone (C), without assuming Maxwell’s light postulate (A), then light would have been treated like any other projectile. We apply the relativity principle in a Newtonian way because all projectiles, including light, add velocities relative to the moving source. However, if Einstein combined relativity (C) with Newtonian kinematics (B), he was led toward emission theory, which eliminates the ether. No ether is needed because light travels directly from the source, much like bullets.

However, Einstein later realized that emission theories led to paradoxes, particularly when attempting to explain simple phenomena like the reflection of light from a moving mirror. If light with velocity $c + v$ struck a mirror perpendicularly, then the reflected light would have velocity $c - v$ according to emission theory, which contradicted experimental observations. Experimentally, the reflected light always travels at the speed of light, regardless of the source or mirror motion. This demonstrates that even the most basic situation, light bouncing off a mirror, contradicts the emission model in a vacuum \cite{Einstein12}. 

The situation becomes even more paradoxical in the media (such as Fizeau’s 1851 water tube experiment) because emission theories predict complete dragging. However, Fizeau observed only partial dragging, perfectly matching Fresnel’s coefficient and Maxwell’s theory, not emission theory.

Consider a tube of length $L$ filled with a dielectric of refractive index $n$, with the fluid moving at speed $v \ll c$ along the tube. A beam is sent \emph{with} the flow and another \emph{against} it; the two paths recombine to
form interference fringes. Let $\lambda_0$ be the vacuum wavelength of the source (frequency $\nu = \dfrac{c}{\lambda_0}$).
In an emission picture, light in the medium at rest travels at $\dfrac{c}{n}$, and when the medium moves at $v$, its speed in the lab is obtained by:
\begin{equation}
    u_\pm^{\mathrm{em}} = \frac{c}{n} \,\pm\, v, \qquad \text{Galilean addition law of velocities,}
\end{equation}
so the times of flight are:%
\footnote{For:
\begin{equation}
\left|\frac{1}{1 \pm \frac{nv}{c}}\right|\ll 1,    
\end{equation}
we use the first-order truncation of the Taylor approximation for small $\varepsilon$:
\begin{equation} \label{epsilon}
\frac{1}{1\pm \varepsilon}\approx 1\mp \varepsilon, \qquad \text{to get:}   
\end{equation}

\begin{equation}
t_\pm^{\mathrm{em}}
\approx
\frac{nL}{c}\!\left(1 \mp \frac{nv}{c}\right).    
\end{equation}}

\begin{equation}
t_\pm^{\mathrm{em}} =\frac{L}{\tfrac{c}{n} \pm v}
= \frac{L}{\frac{c \pm nv}{\,n\,}} 
= \frac{L\,n}{c \pm nv} 
= \frac{nL}{\,c\!\left(1 \pm \frac{nv}{c}\right)} 
= \frac{nL}{c}\,\frac{1}{1 \pm \tfrac{nv}{c}}
\approx \frac{nL}{c}\!\left(1 \mp \frac{nv}{c}\right).
\end{equation}
and the first–order time difference is:
\begin{equation} \label{eq24}
    \Delta t^{\mathrm{em}} \equiv t_-^{\mathrm{em}} - t_+^{\mathrm{em}}
    \approx \boxed{\frac{2Lv}{c^{2}}\, n^{2}.}
\end{equation}
Fresnel’s formula [equation \eqref{eq21} in section \ref{FR}] gives the lab-frame speeds:
\begin{equation}
    u_\pm^{\mathrm{Fr}} = \frac{c}{n} \,\pm\, v\!\left(1 - \frac{1}{n^{2}}\right),  
\end{equation}
Thus:%
\footnote{We start from the left hand-side of \eqref{tFre}:
\begin{equation}
t_\pm^{\mathrm{Fr}}
=
\frac{L}{\tfrac{c}{n} \pm v\!\left(1-\frac{1}{n^2}\right)}. 
\end{equation}
Then we multiply the top and bottom by $\dfrac{n}{c}$ to normalize the denominator by $\dfrac{c}{n}$:
\begin{equation}
t_\pm^{\mathrm{Fr}}
=
\frac{L}{\tfrac{c}{n}\!\left[1 \pm \frac{v n}{c}\!\left(1-\frac{1}{n^2}\right)\right]}
=
\frac{nL}{c}\,
\frac{1}{1 \pm \frac{v n}{c}\!\left(1-\frac{1}{n^2}\right)}.    
\end{equation}
Now we simplify the factor in brackets:
\begin{equation}
\frac{v n}{c}\!\left(1-\frac{1}{n^2}\right)
=
\frac{v n}{c}\cdot \frac{n^2-1}{n^2}
=
\frac{v}{c}\cdot \frac{n^2-1}{n}
=
\frac{v}{c}\left(n-\frac{1}{n}\right). \qquad \text{So:}   
\end{equation}
\begin{equation}
t_\pm^{\mathrm{Fr}}
=
\frac{nL}{c}\,
\frac{1}{1 \pm \frac{v}{c}\!\left(n-\frac{1}{n}\right)}.  \qquad \text{Finally, for:}  
\end{equation}

\begin{equation}
\left|\frac{v}{c}\left(n-\frac{1}{n}\right)\right|\ll 1, \qquad \text{we use \eqref{epsilon} to get:}  \qquad  t_\pm^{\mathrm{Fr}}
\approx
\frac{nL}{c}
\left(
1 \mp \frac{v}{c}\left(n-\frac{1}{n}\right)
\right).    
\end{equation}}

\begin{equation} \label{tFre}
    t_\pm^{\mathrm{Fr}} = \frac{L}{\,\tfrac{c}{n} \pm v(1 - 1/n^{2})\,}
    = \frac{nL}{c}\,\frac{1}{1 \pm \tfrac{v}{c}\!\left(n - \tfrac{1}{n}\right)}
    \approx \frac{nL}{c}\!\left(1 \mp \frac{v}{c}\!\left(n - \tfrac{1}{n}\right)\right),
\end{equation}
and hence:
\begin{equation} \label{eq25}
    \Delta t^{\mathrm{Fr}} \equiv t_-^{\mathrm{Fr}} - t_+^{\mathrm{Fr}}
    \approx \frac{2 n L}{c^{2}}\!\left(n - \frac{1}{n}\right) v
    = \boxed{\frac{2Lv}{c^{2}} (n^{2} - 1).}
\end{equation}
The predictions--equations \eqref{eq24} and \eqref{eq25}--differ by the factor multiplying the common prefactors:
\begin{equation}
    \boxed{ \; n^{2} \; \text{(emission)} \quad \text{vs.} \quad (n^{2} - 1) \; \text{(Fresnel)} \; }.
\end{equation}
But Fizeau observed the \emph{partial-drag} value $(n^{2} - 1)$, not the emission $n^{2}$ value [see equation \eqref{difference} in section \ref{Fize}], in agreement with the Lorentz electron theory and against emission theory.

Einstein further noted that emission theories implied that light could have an arbitrarily large or arbitrarily small velocity, making it challenging to define a plane wave in terms of its intensity, color (or frequency), and polarization state.
If light did not propagate at a universal speed but instead inherited the source velocity (as in emission theories), the entire wave-theoretic framework would collapse. A plane wave presupposes a universal dispersion relation; if the velocity of light $c$ varied with emitter velocity, then the relation $\lambda=\dfrac{c}{v}$ would no longer be fixed. Light of the same frequency emitted by different moving sources would propagate with various wavelengths and phase velocities, so wavefronts could not remain coherent. Interference, diffraction, and the very notion of “monochromatic light” would lose their meaning. In this sense, light ceases to be a wave without a constant propagation speed.

Because of these contradictions, Einstein concluded that it was impossible to construct differential equations and boundary conditions that adequately described light propagation under emission theories. This realization contributed to his rejection of emission theories and his adoption of a new kinematics, in which the speed of light remains invariant in all inertial frames \cite{Einstein12}.

By 1905, Einstein abandoned Newton-Galilean kinematics and adopted a new kinematics in which the speed of light is independent of the motion of its source (A). He retained what emission theory had shown him was possible: physics without an ether. However, light is now understood not as source-dependent particles, but as autonomous fields that propagate through space. This led him to reject the Newtonian addition law of velocities (B) and instead keep (A) and (C), which formed the foundation of special relativity. 

Einstein’s initial attachment to Newtonian principles reflected his careful, stepwise approach to theory building. He did not overthrow established physics all at once; instead, he explored conservative modifications first, encountered contradictions, and only then took the revolutionary step. His intellectual method, testing the limits of existing frameworks before discarding them, was central to his breakthroughs. While this led him to temporary detours, it also ensured that when he finally embraced the radical formulation of special relativity, it was firmly grounded in mathematical consistency and empirical necessity.

In May 1905, Einstein had a rough draft of the relativity paper \cite{CPAE5}, Doc. 28. This draft presented a modification of the theory of space and time. Very likely, the physical definition of simultaneity had already been formulated, and the draft had a purely kinematical part. 
Within the five to six weeks after May 1905, Einstein solved his problem and was able to complete and submit his paper, "Electrodynamics of Moving Bodies," on June 30, 1905, for publication to \emph{Annalen der Physik} \cite{Reiser1930, Wertheimer1916}.

Einstein took the messy tangle of experiments and puzzles --- Fizeau’s partial drag, stellar aberration, ether-drift nulls, Faraday's induction, and the clash between mechanics and electrodynamics --- and created a theory of principle. 
His move was conceptual and operational. He declared that the ether was superfluous and built the whole theory from symmetry and measurement, not from hypotheses about matter deforming in an ether.

\subsection{Clocks, Rods, and Light Signals}

A major misconception about Einstein’s discovery of special relativity, according to Norton, is the belief that his success stemmed from a form of operational thinking—that is, from defining physical concepts solely in terms of measurable quantities such as clocks, rods, and light signals. Norton argues that this interpretation fundamentally misconstrues both the logic and the chronology of Einstein’s work. 

The well-known clock-synchronization thought experiment in Einstein’s 1905 paper, often presented as the key to the relativity of simultaneity, was not the point of discovery but a pedagogical device designed to illustrate results already obtained. Einstein did not arrive at the theory by synchronizing clocks; he used synchronized clocks to explain the theory \cite{Nor1}.

According to Norton, commentators have repeatedly misread the synchronization procedure as if the relativity of simultaneity emerged directly from it — as though Einstein discovered relativity through operational reflections on the exchange of light signals and clock readings. This, Norton insists, reverses the historical and conceptual order. The thought experiment is explanatory, not generative; it clarifies the theory's consequences rather than revealing its source. The so-called “operational thinking” myth confuses the method of exposition with the process of discovery.

Einstein’s real path to special relativity, Norton contends, was the result of years of theoretical struggle within electrodynamics. Through this work, Einstein came to accept two apparently irreconcilable principles: first, the principle of relativity, according to which the laws of physics are the same in all inertial frames; and second, the light postulate, asserting that light propagates at the constant velocity $c$ in every inertial frame. From the reconciliation of these two postulates, the relativity of simultaneity logically followed. The clock-synchronization analysis merely served to make this counterintuitive conclusion intelligible to readers — and perhaps to Einstein himself \cite{Nor1}.

Thus, for Norton, Einstein’s achievement lay not in asking simple, operational questions about measurement, but in recognizing that two seemingly incompatible principles could both be true and in tracing the profound conceptual consequences of that realization.

Even if the synchronization procedure did not play a heuristic role in the genesis of special relativity, it may nonetheless have exerted a reflexive cognitive effect. The explicit formulation of a concept for purposes of exposition can retro-stabilize and sharpen an understanding already in place. In this sense, one might say that Einstein did not arrive at the relativity of simultaneity through the synchronization scheme, but that the act of articulating it may have helped him to see more clearly what he had already grasped implicitly. This nuance does not undercut Norton’s central point: operational reasoning did not generate special relativity. It merely relaxes the sharp dichotomy between “pedagogical” and “internal” by acknowledging the possibility that devices introduced for didactic clarity can, in retrospect, feed back into the agent’s own conceptual self-understanding. The broader moral is methodological: published papers should not be read as diaries of discovery.

\section*{Acknowledgment}

I want to thank the late Prof. John Stachel for spending numerous hours with me discussing Einstein's relativity at the Center for Einstein Studies at Boston University.

\end{document}